\documentclass[12pt,a4paper,twoside]{article}
\usepackage{epsfig}

\newcommand{\be}{\begin{equation}}
\newcommand{\ee}{\end{equation}}
\newcommand{\beqs}{\begin{eqnarray}}
\newcommand{\eeqs}{\end{eqnarray}}
\newcommand{\LL}{{\cal L}}

\newcommand{\tr}{{\rm tr}}

\newcommand{\half}{{1 \over 2}}

\def\NP{{\it Nucl. Phys.\ }}
\def\PL{{\it Phys. Lett.\ }}

\def\PR{{\it Phys. Rev.\ }}
\def\PRL{{\it Phys. Rev. Lett.\ }}
\def\CMP{{\it Comm. Math. Phys.\ }}
\def\JMP{{\it J. Math. Phys.\ }}

\def\JP{{\it J. Phys.\ }}
\def\IJMP{{\it Int. Jour. Mod. Phys.\ }}
\def\MPL{{\it Mod. Phys. Lett. A\ }}
\def\NC{{\it Nuovo Cimento \ }}

\expandafter\ifx\csname mathbbm\endcsname\relax

\else

\fi
\begin{document}
\begin{titlepage}
\begin{flushleft}  
       \hfill                      {\tt hep-th/9902157}\\
       \hfill                      UUITP-1/99\\
       \hfill                       February 1999\\
\end{flushleft}
\vspace*{3mm}
\begin{center}
{\LARGE Generalized statistics in one dimension \\}
\vspace*{8mm}
{\large {\it Les Houches Lectures, Summer 1998} \\}
\vspace*{12mm}
\large Alexios P. Polychronakos\footnote{E-mail:
poly@teorfys.uu.se} \\
\vspace*{5mm}
{\em Institutionen f\"{o}r teoretisk fysik, Box 803 \\
S-751 08  Uppsala, Sweden \/}\\
\vspace*{4mm}
and\\
\vspace*{4mm}
{\em Physics Department, University of Ioannina \\
45110 Ioannina, Greece\/}\\
\vspace*{15mm}
\end{center}

\begin{abstract}
An exposition of the different definitions and approaches to
quantum statistics is given, with emphasis in one-dimensional
situations. Permutation statistics, scattering statistics and
exclusion statistics are analyzed. The Calogero model, matrix
model and spin chain models constitute specific realizations.

\end{abstract}

\end{titlepage}

\section{Introduction}

Quantum statistics, as the name suggests, is the way that the
quantum mechanical properties of particles influence their
statistical mechanics. Two features of quantum mechanics are crucial
for the difference between classical and quantum statistics: the
indistinguishability of quantum particles and the `finite room' that
exists in a given domain of the phase space of a quantum system.

The first feature can be incorporated `by hand' into classical
statistical mechanics (although there is no underlying principle
for doing so) and this will resolve the Gibbs paradox, although
by itself it would change little else. The second feature can
likewise be incorporated in an ad hoc way, which will assign a unique
value to the additive constant of the entropy of a classical system,
but again without further physical impact. It is the combination of
the two that produces nontrivial results, especially in the domain
where quantum phenomena prevail, that is, high densities or low
temperatures.

The statistics of observed particles seem to be exhausted by the
two well-known cases of Bose and Fermi. In this sense, the quest for
more exotic possibilities seems to be a bit academic or even quixotic.
It is, nevertheless, of interest to explore the possibilities as a
way of obtaining a more lucid {\it effective} description of the dynamics
of (otherwise normal) {\it interacting} particles. In this sense, we are
trading one kind of intuition for another, hoping to make a profit
in the bargain. 

There are several approaches to defining or implementing quantum
statistics. They are roughly classified into `fundamental' (group theoretic,
operator etc.) and `phenomenological' (statistical). While exploring them we 
must keep in mind that, at the end of the day, we should provide a 
realization, that is, a system which is (more or less) physical and
exhibits these statistics through its dynamics. The less the number of
space dimensions, the more easy it is to find such realizations. 
The concept itself of statistics, actually, depends crucially on the
dimensionality of space. Topological considerations allow for generalized
abelian statistics in 2+1 dimensions \cite{LM,GMS,Wil,WuA}. 
Such particles, called anyons, have
found use in the physics of the quantum Hall effect and (probaby)
high-temperature superconductivity. There are excellent reviews on 
anyons \cite{Jan} and we will not deal with them here.

In these lectures the bulk of the material will concern one-dimensional
situations, although some considerations in principle applicable to any
dimension will also be included. The purpose is to present enough
of the concepts and material on the subject to spur the interest of
the reader for further study. You are warned that all formulae
in these lectures are simply mnemonics of the correct results: mistakes
abound! (The written version is only marginally better.) One standard 
exercise proposed here is to rederive all results, fixing the signs and
factors in the process. Hopefully substance survived the sloppyness.

So, right to the task.

\section{Permutation group approach}

As a way of introduction and covering standard ground,
we will study the context in which statistics initially arose, namely
the permutation group in quantum mechanics \cite{GR}-\cite{GGG}. 
The main emphasis will be
in connecting to the path integral and pointing the possibilities
for generalizations within this approach.

The main themes of this lecture are:

\noindent
a) Eliminating the `gauge' multiplicity of states originating from
particle permutations. This will, in general, lead to the appearance
of internal degrees of freedom that are the residue of
`identities' (distinguishability) of particles.

\noindent
b) Implementing the path integral for identical particles. Their
indistinguishability calls for including paths that lead to any
permutation of the final position. Determining the right weights of
these sectors (which are not necessarily phases) will fix the statistics.

\subsection{Realization of the reduced Hilbert space}

We will work in the first-quantized picture, in which the number of
particles is fixed but the particles are nevertheless taken to be
{\it indistinguishable}. This is to be contrasted to the weaker statement
that the particles are merely {\it identical}. To fix the ideas, let us
call $\{ x_i \}$ the coordinates of these particles, with $i=1 \dots N$
enumerating the particles, and assume, for now, that there are no
internal degrees of freedom. The $x_i$ can be in a space of any dimension.
The Hilbert space of this system is spanned by the position 
eigenstates $|x_1 , \dots x_N > \equiv | x >$. 
Let us also define the operators $P \in S_N$ 
that permute the quantum numbers of the particles. There are $N!$
such operators forming the permutation (or symmetric) group of $N$
objects $S_N$.
The collection of the above states for a set of distinct $x_i$
transforms in the $N!$-dimensional regular (defining) representation
of $S_N$
\be
P |x> \equiv |Px> = |x_{P(1)}, \dots x_{P(N)} >
\ee
Notice that if any of the coordinates $x_i$ coincide the above is not the 
full defining representation any more. The set of such states, however, is
of measure zero (the coordinate space assumed continuous) and thus they can
be safely ignored. (We assume that there are no interactions singular at 
coincidence points that might dynamically make such states of nonzero weight.)

If the particles are identical then the hamiltonian treats them all
equally and thus we have
\be
[H,P]=0
\ee
for any $P$. If however, the particles are also indistinguishable,
this means that there is {\it no physical way} to ever be able to
tell the particles from each other. Thus, not only the hamiltonian,
but all conceivable perturbations of this hamiltonian must commute with
the permutation operators. (These perturbations would correspond to
various devices, dies and other tricks that we would attach to the system
in our effort to tell the particles from each other.) In short, the
permutation operators must commute with {\it all physical observables} $A$:
\be
[A,P]=0
\ee
When such an operator $P$ commutes with all (physical) operators in a 
system we have a superselection rule: there is no possible transition or 
interference between states corresponding to different eigenvalues of
this operator and thus we can choose a subspace with a fixed eigenvalue
and stay with it for ever. (This is also called `picking a theta-vacuum'
in particle gauge theories.) In our case, there is a whole set of 
operators commuting with everything else (the group $S_N$) and thus the
generalized `theta-vacuum' states consist of {\it irreducible representations}
(irreps) of $S_N$. Further, states that are connected 
to each other through the action of operators $P$ cannot be distinguished
by any physical operator and therefore must be identified as a unique
physical state. In other words, the permutation group must be viewed
as a discrete {\it gauge group} producing unphysical transformations,
that is, copies of the same physical system. We must, therefore:

$\bullet$ Project on a subspace corresponding to a particular irrep
$R$ of $S_N$

$\bullet$ Identify elements within this subspace grouping in the same
irrep as a unique physical state.

\noindent
The above procedure and choice of $R$ corresponds to a choice of statistics.

We will implement this
in the coordinate representation $|x_1 , \dots x_N > \equiv | x >$. 
Projecting the Hilbert space to an irrep $R$ of $S_N$ amounts to keeping
only linear combinations of states within this multiplet that transform in 
$R$, that is,
\be
|a;x> = \sum_P C_a (P) P |x> ~,~~~ a = 1, \dots d_R ~,~~d_R =dim(R).
\ee
where the sum is over all elements of the permutation group and $C_a (P)$ 
are appropriately chosen coefficients. If we denote with $R_{ab} (P)$
the matrix elements of the permutation $P$ in the representations $R$,
\be
P |a,x> = \sum_b R_{ab} (P^{-1} ) |b,x>
\label{Ca}
\ee
(The appearance of $P^{-1}$ above is necessary so that successive
transformations combine in the right order.)

The defining representation decomposes into irreducible components,
classified by Young tableaux, each appearing with a certain multiplicity.
Should we keep only one irrep out of each multiplicity or the whole multiplet?
To decide it, note that if instead of the base state $|x>$ for the
construction of the states $|a,x>$ we choose a different permutation
$P_o |x>$, then although the new states $|a, P_o x>$ constructed through
(\ref{Ca}) still transform in the irrep $R$, in general they are {\it not}
linear combinations of $|a,x>$ but rather span a different copy of $R$.
Since we can continuously move in the configuration space from $|x>$ to
$P_o |x>$, we conclude that we must keep {\it all} irreps $R$ within each
multiplet. (In other words, although for each point in the Hilbert space
$|x>$ this multiplet is reducible, the fiber of these representations
over the Hilbert space is connected and irreducible.)

To realize explicitly the above, we construct the states
\be
|ab,x> = \sqrt{d_R \over N!} \sum_P R_{ab} (P) P |x>
\label{ab}
\ee
Using the group property of the representation $R(P_1 ) R(P_2) = R(P_1 P_2 )$,
we deduce that under the action of the group $S_N$ and under change of
base point $x$ the above states transform as:
\be
P |ab,x> = \sum_c R_{ac} (P^{-1} ) |cb,x>~,~~~
|ab,Px> = \sum_c R_{cb} (P^{-1} ) |ac,x>
\label{LR}
\ee
Thus we see that the first index in these states labels the different
elements of a single irrep $R$, while the second index labels the
different equivalent irreps in the multiplet. Since both indices take
$d_R$ values, we recover the standard result that each irrep of $S_N$
is embedded in the defining representation a number of times equal to
its dimension.

Consider now the matrix element $<ab,x| A |cd,y>$, where $A$ is any
physical operator, that is, any operator commuting with all elements $P$
of $S_N$. Substituting the definition (\ref{ab}) and using the unitarity
of $P$ ($P^\dagger = P^{-1}$) and of $R$ ($R_{ab}^* (P) = R_{ba} (P^{-1} )$)
we obtain, after a change in summation variable,
\be
<ab,x| A |cd,y> = {d_R \over N!} \sum_{P,P',e} R_{be} (P' ) R_{ea} ( P^{-1} )
R_{cd} (P) <x|A P'|y>
\ee
Using further the orthogonality (Shur's) relation (see, e.g., \cite{HAM})
\be
\sum_P R_{ab} (P) R_{cd} (P^{-1} ) = {N! \over d_R} \delta_{ad} \delta_{bc}
\ee
we finally obtain
\be
<ab,x| A |cd,y> = \sum_P \delta_{ac} R_{bd} (P) <x|A|Py>
\label{AB}
\ee
It is clear from (\ref{AB}) that there is no possible transition between
states with different first index. Further, for states with the same
index, the amplitude is independent of the index. Thus, the first index 
$a$ in the state $|ab,x>$ propagates trivially. Since this is the index 
that corresponds to the
different but physically equivalent states within each irrep $R$, we
conclude that the required projection of the Hilbert space to the physical 
subspace amounts to simply omitting this index from all states. (That is,
freeze this index to the same fixed value for all states of the theory;
no physical quantity will ever depend on the choice of this value.) On
the other hand, the second index, corresponding to different equivalent
irreps, does {\it not} propagate trivially and must, as argued before,
be kept. We are led therefore to the physical states $|ba,x> \to |a,x>$. 

Let us first choose $A=1$. Then (\ref{AB}) provides the overlap between
the states
\be
<ab,x|cd,y> = \sum_P \delta_{ac} R_{bd} (P) \delta (x - Py )
\label{Norm}
\ee
For $x$ in the neighborhood of $y$ it is $P=1$ which contributes to
the normalization, for which $R_{bd} (1) = \delta_{bd}$ and we recover
the standard continuous normalization between the states. 

Choose, now, $A=e^{-i H t}$, where $H$ is the hamiltonian, and thus
find the propagator $G(ab,x;cd,y|t)$ between the states of the system. 
Projecting to the physical states $|ba,x> \to |a,x>$ we obtain
the propagator
\be
G_R (a,x;b,y|t) = \sum_P R_{ab} (P)\, G(x,Py;t)
\label{G}
\ee
where $G(x,Py;t) =~ <x| e^{-iHt} P|y>$ is the usual many-body propagator.
We note that, due to the transformation property (\ref{LR}), the states
$|a,Px>$ are linear combinations of states $|a,x>$. Therefore, projecting
down to the physical subspace corresponding to $R$ amounts to trading
the original $N!$ copies of physically equivalent states $|Px>$ for
a number $d_R$ of {\it global internal degrees of freedom} for the system,
labeled by the index $a$. 

\subsection{Path integral and generalized statistics}

It is now easy to write down the path integral corresponding to identical
particles quantized in the $R$-irrep of $S_N$. $G(x,Py;t)$ can be expressed
as an $N$-body path integral in the standard way, with particles starting 
from positions $x_i$ and ending in positions $Py_i = y_{P(i)}$.
Since all permutations of particle
positions are physically equivalent, (\ref{G}) instructs 
us to sum over {\it all}
sectors where particles end up in such permuted positions, weighted with the
factors $R_{ab} (P)$ depending on the internal degrees of freedom of the 
initial and final states. From (\ref{ab}, \ref{Norm}) 
we can write the completeness relation within the physical subspace
\be
I_R = \int {d^N x \over N!} \sum_a |a,x> <a,x|
\label{Comp}
\ee
and with the use of (\ref{Comp}) 
it is easy to prove that the above path integral
is unitary, that is, 
\be
\int {d^N y \over N!} \sum_b G(a,x;b,y|t) \, G(b,y;c,z|t') 
= G(a,x;c,z|t+t') 
\ee

We can extend this construction to a more general class of statistics,
which includes the cases of the so-called `parabosons' and `parafermions' 
as special cases. The generalization consists in allowing more than
one irrep to appear in the Hilbert space, and keeping more than one state
in each irrep. This seems unmotivated and against the spirit of the
reduction by the full superselection rule, but it is consistent.
It could mean, for instance, that the particles have some hidden internal
degrees of freedom accounting for the extra degeneracy, which are invisible
to the present hamiltonian but may become dynamically relevant later. (This,
then, would augment the space of physical operators in the theory.)
The most general situation is that we include $C_R$ states from each irrep
$R$. The statistics is fixed by the set of non-negative integers $C_R$.

The generalization of the results to this more general
situation is straightforward. The internal degree 
of freedom now takes values $A=(R,a,\alpha)$, where $a=1, \dots d_R$ 
labels the inequivalent $R$-irreps and and $\alpha=1, \dots C_R$ labels
the states kept within each irrep. $(a,\alpha )$ constitute the internal 
degrees of freedom within each irrep. So, overall, $A$ takes 
$\sum_R C_R d_R$ different values.
The propagator (and corresponding path integral) is obviously
\be
G_S (A,x;B,y|t) = \sum_P S(P)_{AB} \, G(x,Py;t) 
\ee
where
\be
S(P)_{AB} = \delta_{R_A , R_B } ~\delta_{\alpha \beta} ~({R_A})_{ab} (P)
\ee
The definition of parabosons (parafermions) of order $p$ is that we
include once all irreps
with up to $p$ rows (columns) in their Young tableaux. We have, 
therefore, $C_R =1$ for such irreps and $C_R =0$ for the rest.
We note that the irreps for parafermions are 
the duals of those for parabosons (the dual of a tableau is the tableau with
rows and columns interchanged). In an appropriate basis, the representation
matrices of dual irreps $R, {\tilde R}$ are real and satisfy
\be
{\tilde R}_{ab} (P) = (-1)^P R_{ab} (P)
\ee
where $(-1)^P$ is the parity of the permutation. We arrive then at the 
relation between the weights for parabosons and parafermions of order $p$:
\be
S_{pF} (P)_{AB} = (-1)^P S_{pB} (P)_{AB} 
\ee
This extends a similar relation for ordinary fermions and bosons \cite{LDW}, 
for which there are no internal degrees of freedom and $S_B (P) =1$.

{}From the path integral we can evaluate the partition function, by simply
shifting to the euclidean periodic propagator $G_E (\beta) = e^{-\beta H}$ 
and summing over all initial and final states, with the measure implied by 
(\ref{Comp}). Given that
\be
\sum_a R_{aa} (P) = \tr R (P) = \chi_R (P)
\ee
we get the expression in terms of the characters of $S_N$
\be
Z_S (T) = \int {d^N x \over N!} \sum_P S(P) <x| G_E (\beta) |Px>
\ee
where
\be
S(P) = \sum_R C_R ~\chi_R (P) 
\label{ZS}
\ee
We also note the inversion formula
\be
C_R = {1\over N!} \sum_P S(P) \chi_R (P)
\ee
which allows to recover $C_R$ if we know $S(P)$ for all permutations.
The interpretation in terms of a periodic euclidean path integral
is obvious. The characters $\chi_R (P)$ are a set of integers, and thus
the ``statistical factors" $S(P)$ weighing each topological sector of the
path integral are (positive or negative) integers. In the case of
parabosons of any order $p$, however, we note that the statistical 
weights are {\it positive} (or zero) integers. The ones for parafermions
can be either positive or negative, as given by 
\be
S_{pF} (P) = (-1)^P S_{pB} (P) ~,~~~ S_{pB} (P) \ge 0
\ee
We do not have a general formula for $S_{pB} (P)$ for arbitrary $p$.
For non-interacting particles the partition function can be expressed
in terms of the characters of $SU(N)$ (Shur's functions) \cite{SU,CHA}.

We note here that the case of distinguishable particles 
(``infinite statistics" \cite{DHR,GOV,GRB}),
is also included in this formalism, by accepting all states in each irrep,
that is, $C_R = d_R$. Since $R$ appears exactly $d_R$ times in the defining 
representation of $S_N$, $S(P)$ above becomes the trace of $P$ in the
defining representation. But all $P\neq 1$ are off-diagonal in the defining
representation, so we get $S_{inf} (P) = N! \, \delta_{P,1}$, recovering
\be
Z (T) = \int d^N x <x| G_E (\beta) |x>
\ee
for distinguishable particles.

\subsection{Cluster decomposition and factorizability}

Parastatistics particle obey the cluster
decomposition principle, in the sense that the density matrix obtained
by tracing over a subset of particles which decouple from the system
can be constructed as a possible density matrix of the reduced system
of remaining particles \cite{LS,HST}. 
{}From (\ref{ZS}), however, we see that the partition
function of two dynamically isolated sets of particles $N_1$ and $N_2$
does {\it not} factorize into the product of the two partition functions,
since the statistical weights $S(P)$ in general do not factorize into
$S(P_1 ) S(P_2 )$ when $P$ is the product of two commuting elements
$P_1$ and $P_2$. Equivalently, this means that the occupation degeneracy
$D(p_1 , p_2 , \dots )$ of $p_1 , p_2 , \dots$ particles occupying 
a set of levels $\epsilon_1 , \epsilon_2 , \dots$ 
does not factorize into the product of individual occupation
degeneracies for each level $\epsilon_i$. There is an effective
`coupling' between the particles. This has important physical
implications. If the two sets of particles are totally isolated, it
does not make sense to evaluate the partition function of the total
system, since the statistical distribution can never relax to the one
predicted by that partition function. The individual partition functions
of the subsystems are the relevant ones. If, however, the two sets are
only weakly coupled, then initially each set will distribute according
to its reduced partition function, but after some relaxation time (depending
on the strength of the coupling between the two sets) they will relax to
the joint distribution function, which, we stress, will not even
approximately equal the product of the individual ones. Thus, cluster
decomposition holds in an absolute sense but fails in a
more realistic sense. In contrast, fermions and bosons
respect cluster decomposition in both senses.

We summarize by pointing out that the most general statistics of the
type examined here is parametrized by any of three possible sets of numbers.
The first is, as just stated, the number of states $C_R$ accepted for
each irrep $R$ of $S_N$. The second is the
statistical weights $S(P)$ appearing in the partition function (euclidean
path integral). These weights are invariant under conjugation
of $P \to Q P Q^{-1}$.  Finally, we could use the degeneracy of a many-body
occupancy state, call it $D[p_i ]$, as our definition. (This was the 
starting point of the first known attempt to generalized statistics,
the `Gentile statistics.' \cite{GEN}) These three sets contain as many
elements as the partitions of $N$ and are all equivalent.

What are the restrictions or criteria to be imposed on the above parameters?
The first one is unitarity, that is, the existence of a well-defined Hilbert
space with positive metric. This requires that $C_R$ be non-negative (no 
negative norm states) integers (no ``fractional dimension" states). The other
will be what we call ``strong cluster decomposition principle" 
or ``factorizability,'' that the 
partition function of isolated systems factorizes. This is a physical 
criterion, rather than a consistency requirement. To summarize:

$\bullet$ Unitarity: $C_R$ non-negative integers \hfill

$\bullet$ Factorizability: $S(P) = \prod_{n \in cycles(P)} S(n) 
~~~{\rm or}~~~D[p_i ] = \prod_i D( p_i )$ \hfill

\noindent
Factorizability, in particular, implies that the grand partition function
in the case of noninteracting particles will be obtained
by exponentiating the sum of all {\it connected} path integrals ($P$ a 
cyclic permutation of degree $n$) with weights $S(n)/n$ ($1/n$ is the 
symmetry factor of this path integral, corresponding to cyclic relabelings
of the particles). The grand partition function will factorize
into a product of partition functions for each level $\epsilon_i$. Thus,
$S(n)$ are cluster coefficients connected to $D(n)$ in the standard way 
\be
\sum_{p=0}^\infty D(p) z^p = 
\exp \left( \sum_{n=1}^\infty {S(n) \over n} z^n \right)
\ee
The above formula, in fact, provides the easiest way to relate 
$D[p_i ]$ and $S(P)$ in the general case (no strong cluster property)
but we will not enter into this.

If we assume that $S(1) = D(1) = 1$, then it is easy to verify that the
{\it only} solution of the above two criteria is ordinary fermions and
bosons. The situation is different, however, when $S(1) = D(1) = q >1$ (this
would mean, e.g., that the particles come a priori in $q$ different 
``flavors"). The possibilities are manifold. All these generalized statistics
share the following generic features:

$\bullet$ The degeneracy of the state where $n$ particles occupy different
levels is $q^n$. (Indeed, $D(1,1,\dots 1) = D(1)^n = q^n$.)

$\bullet$ If state $A$ can be obtained from state $B$ by `lumping'
together particles that previously occupied different levels, then
$D(A) \le D(B)$. (E.g., $D(3) \le D(2,1) \le D(1,1,1)$.)

The above possibilities include the obvious special cases of $q_1$
bosonic flavors and $q_2$ fermionic ones ($q_1 + q_2 = q$), for which
$S(n)=q_1 - (-1)^n q_2$, along with many other. As an example, we give
the first few degeneracies for many-particle level occupation for all 
statistics with $q=2$:

$D(1)=2, ~~D(2)=4, ~~D(3)=8$

$D(1)=2, ~~D(2)=3, ~~D(3)=6,5,4(B+B)$

$D(1)=2, ~~D(2)=2, ~~D(3)=4,3,2(B+F),1,0$

$D(1)=2, ~~D(2)=1, ~~D(3)=2,1,0(F+F)$

$D(1)=2, ~~D(2)=0, ~~D(3)=0$

The specific choices denoted by $B+B$, $B+F$ and $F+F$ are the ones
corresponding to two bosonic, one bosonic and one fermionic, and two
fermionic flavors respectively. The topmost statistics could be termed
``superbosons" and the bottom one ``superfermions" of order 2. We also
remark here that the ``$(p,q)$-statistics'' introduced in \cite{HST}
can be realized as particles with $p$ bosonic and $q$ fermionic
flavors, where we identify each multiplet transforming irreducibly under
the supergroup $SU(p,q)$ as a unique physical state.

Finally, we comment on `Gentile statistics' \cite{GEN}. 
The rule is simply that up to $p$ particles can be put in
each single-particle level. This corresponds to $D(n)=1$ for $n \le p$,
and $D(n)=0$ otherwise. This has been criticized \cite{GGG} 
on the grounds that
fixing the allowed occupations for each single-particle state is not
a statement invariant under change of single-particle basis. It should
be clear from this lecture that any statistics satisfying the
unitarity requirement is consistent and basis-independent. Therefore,
Gentile statistics must violate unitarity. Indeed, it is easy to check
that all weights $C_R$ for such statistics are integers (this is
generic for all statistics with integer $D(n)$), but not necessarily
positive. In the specific case of $p=2$, e.g., where up to double
occupancy of each level is allowed, the degeneracies of each irrep of 
$S_N$ (parametrized, as usual, by the length of Young tableau rows) 
up to $N=5$ are
\be
C_2 = C_{21} = C_{22} = C_{221} =1,~ C_{111} = C_{1111} = C_{2111} =-1,
~~{\rm else}~C_R =0
\ee
We see that representations $111$, $1111$, $2111$ correspond to
{\it ghost} (negative norm) states and their effect is to subtract
(rather than add) degrees of freedom. 

In conclusion, we see that the permutation group approach to
generalizing statistics gives results of rather limited interest.
The resulting statistics can be though of as particles with internal 
degrees of freedom, with perhaps some superselection rule that
forbids the manifestation of all possible internal states.
To find something more exotic we must consider other approaches.

\section{One-dimensional systems: Calogero model}

Let us now specialize in one dimension and see in what different
angle we can approach the problem of statistics. The peculiarity
of a lineal world is that, kinematically, particles cannot
exchange positions without ``bumping" onto each other. This has
tangible physical consequences and makes the notion of statistics
in one dimension rather special. Let's see:

A. Since configurations related by particle permutation are
`gauge' copies of each other, we could `gauge fix' and restrict
ourselves to only one gauge copy. In one dimension the configuration
space of $N$ indistinguishable particles breaks into $N!$ sectors
classified by the ordering of the coordinates: $x_1 < \dots < x_N$
and its permutations. Restricting to one sector we are faced with the
problem of boundary conditions on the wavefunction at
the boundary of this space, when
two or more coordinates become equal, so that we preserve hermiticity.
The choice of boundary conditions can be interpreted as a choice of
statistics. We mention two possibilities:

a) A linear local boundary condition:
\be
\left. \psi + \lambda \partial_n \psi \right|_{\rm boundary} =0
\ee
where $\partial_n$ is the normal derivative at the boundary \cite{PolA}.
Clearly $\lambda =0$ (Neumann) corresponds to fermions; we can analytically
extend $\psi$ in the other sectors in an antisymmetric way. Similarly,
$\lambda=\infty$ (Dirichlet) corresponds to bosons. Any other choice
would be some intermediate statistics. $\lambda$, however, introduces
a length scale and is not a very satisfactory statistics definition.
At any rate, this system is equivalent to bosons with a delta-function
two-body interaction of strength proportional to $\lambda^{-1}$.

b) Fix the analytical behavior of $\psi$ at the boundary as
\be
\psi \sim x^\ell ~~{\rm as}~~ x \to 0
\label{bcell}
\ee
This way the probability current would scale as $\psi \partial_n \psi
\sim \ell x^{2\ell-1}$ and would vanish if $\ell >\half$ or $\ell=0$.
This would define the statistics through the dimensionless parameter $\ell$.
Notice that this behavior would require a two-body potential behaving
like $\frac{\ell(\ell-1)}{x^2}$ near coincidence points. This is the
first glimpse at the inverse-square potential arising in the statistics
context.

B. We could, alternatively, examine the scattering phase $\theta$ between
particles. At high relative energies the two-body scattering phase
approaches the values
\be
\theta \to 0 ~~{\rm for~bosons},~~~
\theta \to \pi ~~{\rm for~fermions}
\ee
irrespective of the details of their interaction (provided it is not
too singular). The corresponding many-body phase is the sum of two-body
ones. So, if we see a system where this phase goes to the value
$\ell \pi$ we can interpret it as one with generalized statistics
of order $\ell$ \cite{PolA}. 

As we shall see, this definition also leads to the inverse-square potential.
There are hints from algebraic considerations also pointing to this type 
of potential \cite{LMIS,IL,ILMPV}.
This is enough motivation, at any rate, to examine particles with
that type of two-body interaction potentials. This is known as the
Calogero model, or, more completely, as the Calogero-Sutherland-Moser
(CSM) model \cite{Calo,Suth,Mos}.

\subsection{The Calogero-Sutherland-Moser model}

This is a system of nonrelativistic identical particles on the line
with pairwise inverse-square interactions. The basic hamiltonian is
\be
H = \sum_{i=1}^N \half p_i^2 + \sum_{i<j} \frac{g}{(x_i - x_j )^2}
\ee
This is the `free' (scattering) Calogero system on the line \cite{Calo}. The
particle masses $m$ have been scaled to unity. This is the only
scale-free two-body potential that one can have (the potential,
quantum mechanically, scales like the kinetic term). An external
harmonic oscillator potential can also be added to confine the system
without spoiling its features. This is the harmonic Calogero model.
Alternatively, one could consider a
periodic version of the system. The particles now interact through
infinitely many periodic copies of themselves and the two-body potential
becomes 
\be
V(x) = \sum_{n=-\infty}^\infty \frac{g}{(x+2\pi n)^2}
= \frac{g}{\left(2 \sin\frac{x}{2} \right)^2}
\ee
(we scaled the length of the box to $2\pi$). This is the Sutherland
model \cite{Suth}. There are other versions of this class of models
that will not concern us here. The classic report \cite{OP} covers
these systems in detail. See also \cite{HP} for recent mathematical work.

Classically the coupling constant $g$ should be positive to ensure
particles are not `sucked' into each other. Quantum mechanically
the uncertainty principle works in our favor and the minimum allowed
value for $g$ is $g= -\frac{1}{4}$ (we put henceforth $\hbar =1$).
For later convenience, it is useful to parametrize $g$ in the fashion
\be
g = \ell(\ell-1)
\ee
in which case the minimum value is naturally obtained for $\ell = \half$.

The above system is integrable, which means that there are $N$
integrals of motion in convolution, that is, $N$ functions on
phase space with vanishing Poisson brackets:
\be
\{ I_n , I_m \} =0 ~,~~~ n,m=1 \dots N
\ee
For the scattering system (no external potential) $I_1$ is the total
momentum, $I_2$ is the total energy amd the higher $I_n$ are higher
polynomials in the momenta also involving the two-body potentials.
We will not belabor here their form nor prove integrability at this
point, since this is beyond our scope and will, at any rate, be
dealt with in later lectures. We will only state without proof the
qualitative features relevant to our purposes. 

The key interesting property of the above model, that
sets it apart from other merely integrable models, is that, both 
classically and quantum mechanically, it mimics as closely as 
possible a system of free particles. Let us first look at its
classical behavior. The motion is a scattering event.
Asymptotically, at times $t=\pm \infty$,
the particles are far away, the potentials drop off to zero and 
motion is free. When they come together, of course, they interact and
steer away from their straight paths. Interestingly, however, when
they are done interacting, they resume their previous free paths
as if nothing happened. Not only are their asymptotic momenta the
same as before scattering, but also the asymptotic positions (scattering
parameters) are the same. There is no time delay of the particles
at the scattering region. The only effect is an overall reshuffling
of the particles. Thus, if one cannot tell particles from each other,
and if one only looks at scattering properties, the system looks free!

This behavior carries over to quantum mechanics. The asymptotic
scattering momenta are the same before and after scattering. The fact
that there is, further, no time delay translates into the fact that the 
scattering phase shift is independent of the momenta. Thus it can only
be a function of the coupling constant and the total number of
particles. It is, actually, a very suggestive function:
\be
\theta_{sc} = \frac{N(N-1)}{2} \ell \pi
\ee
Thus the phase is simply $\ell \pi$ times the total number of
two-body exchanges that would occur in the scattering of free particles.
Clearly the case $\ell =0$ would correspond to free bosons and $\ell
=1$ to free fermions (for these two values the potential vanishes
and the system is, indeed, free). For any other value we can interpret
this system as free particles with generalized statistics $\ell$.

A word on the permutation properties of this system is in order.
The inverse-square potential is quantum mechanically impenetrable, and
thus the `ordinary' statistics of the particles (symmetry of the
wavefunction) is irrelevant: if the particles are in one of the $N!$
ordering sectors they will stay there for ever. The wavefunction 
could be extended to the other sectors in a symmetric, antisymmetric
or any other way, but this is irrelevant for physics. No interference
between the sectors will ever take place. All states have a trivial
$N!$ degeneracy. (That is to say all states in all irreps of $S_N$
have the same physical properties.) Permutation statistics are 
therefore irrelevant and we can safely talk about the effective 
statistics as produced by their coupling constant $\ell$.

The behavior of the wavefunction near coincidence points in as in
(\ref{bcell}). This system, thus, satisfies both our boundary condition
and scattering phase criteria for generalized statistics.

Let us also review the properties of the confined systems. In the presence
of an external harmonic potential of the form
\be
V = \sum_i \half \omega^2 x_i^2
\ee
the energy spectrum of a system of uncoupled particles would be
\be
E = \frac{N}{2} \omega + \sum_i n_i \omega
\label{Efree}
\ee
The $n_i$ are nonnegative integers satisfying
\be
n_1 \le \dots \le n_N ~~~{\rm for~bosons}
\ee
\be
n_1 < \dots < n_N ~~~{\rm for~fermions}
\ee
The actual spectrum of this model is
\be
E = \frac{N}{2} \omega + \ell \frac{N(N-1)}{2} \omega + \sum_i n_i \omega
\ee
with $n_i$ being `pseudo-excitation numbers' obeying bosonic selection rules:
$n_i \le n_{i+1}$.
Defining the `quasi-excitation numbers'
\be
{\bar n}_i = n_i + (i-1) \ell
\ee
we can check that the expression of the spectrum in terms of the ${\bar n}_i$
is identical to the free one (\ref{Efree}) but with the quantum numbers
now obeying the selection rule
\be
{\bar n}_i \le {\bar n}_{i+1} - \ell
\label{nselect}
\ee
This is a sort of exclusion principle that requires the particle quantum
numbers to be at least a distance $\ell$ apart (as contrasted to 1 for
fermions and 0 for bosons). Again, a generalized statistics interpretation
is manifest \cite{Isa}. 

Let us clarify that the above numbers ${\bar n}_i$ are no more integers.
They do, however, increase in integer increments. The rule is that
the ground state is determined by the minimal allowed nonnegative
values for ${\bar n}_i$ obeying (\ref{nselect}) while the excited states
are obtained by all integer increments of these values that still
obey (\ref{nselect}).

The periodic (Sutherland) model has similar properties. Its spectrum is
\be
E = \sum_i \half k_i^2 + \ell \sum_{i<j} (k_j - k_i ) +
\ell^2 \frac{N(N^2 -1)}{24}
\ee
with the `pseudo-momenta' $k_i$ being integers satisfying bosonic
rules: $k_i \le k_{i+1}$. This looks rather different than the corresponding
free expression (for $\ell=0$). Defining, however, again `quasi-momenta'
\be
p_i = k_i + \ell \left( i - \frac{N+1}{2} \right)
\ee
we can check that the expression for the spectrum becomes
\be
E = \sum_i \half p_i^2 
\ee
that is, the free expression. The quasi-momenta satisfy
\be
p_i \le p_{i+1} -\ell
\label{pselect}
\ee
that is, the same selection rule as the ${\bar n}_i$ before. Again,
we observe a generalization of the fermionic and bosonic selection
rules corresponding to statistics $\ell$. The ground state is the
minimal (nearest to zero) numbers satisfying (\ref{pselect}) while
excitations correspond to integer increments (or decrements) of
the $p_i$ still satisfying (\ref{pselect}).

Let us note that the above rule for $\ell =1$ reproduces the 
fermionic spectrum of particles with periodic boundary conditions
for odd $N$ and {\it anti}-periodic ones for even $N$: in the 
odd (even) $N$ case the momenta are quantized to (half-) integers.
This has a natural interpretation: when we take a particle around
the circle it goes over $N-1$ other particles. If we require the
phase shift of the wavefunction in this process to agree with the
minus signs picked up from the $N-1$ fermion exchanges we recover
the previous rule. We stress that, for free particles, this is
not a consistency requirement but rather an aesthetic rule. At any
rate, this is what the Sutherland model chooses to do!

In conclusion we see that the CSM model can be though of as a
system of particles obeying generalized statistics. This manifests
in terms of the boundary conditions of the wavefunction, the scattering
phases and, most significantly, through a peculiar `level repulsion'
of their quantum numbers generalizing the Fermi exclusion principle.

\subsection{Large-N properties of the CSM model and duality}

Let us examine, now, the properties of the CSM model as the
number of particles grows large. At zero temperature, a 
non-interacting
fermion system would form a Fermi sea. The corresponding
`Fermi surface' in one dimension degenerates to points. For
the system in an external harmonic potential there is just
one point corresponding to the highest excitation $n_F = N-1$.
For the free periodic system we would have two Fermi momenta at
$\pm p_F = \pm \frac{N-1}{2}$. Excitations over this ground state are,
then, conveniently classified in terms of particles (a filled Fermi
sea with an isolated particle above or below) and holes (a filled
sea with one unoccupied state inside it). 

Interestingly, the CSM model presents a similar picture. The
qualitative features of both the Calogero and the Sutherland model
are similar, so we pick the latter as most closely representing a
gas of free particles in a box. From (\ref{pselect}) above we
see that the ground state also forms a `quasi-Fermi sea' (or should
we call it a `Luttinger sea'?) with Fermi levels rescaled by a
factor $\ell$: $p_F = \ell \frac{N-1}{2}$. Its minimal excitations
are analogous to the ones of a Fermi sea, but not quite:

$\bullet$ A particle would be an isolated occupied quasimomentum
above or below a completely filled sea of quasimomenta. Particles

\noindent
--are excited in units of 1 (the increments of their quasimomentum).

\noindent
--take up a space $\ell$ in quasimomentum (since they cannot be
`packed' closer than $\ell$ units apart).

$\bullet$ A hole would be an isolated empty space inside an
otherwise occupied sea. Interestingly, the minimal such excitation
is {\it not} obtained by removing one particle from the sea, but
rather by incrementing all quasimomenta of the sea above the place
where we want to create the hole by one unit. Holes

\noindent
--are excited in units of $\ell$. Indeed, since the distance of
quasimomenta in the sea is $\ell$, the possible positions of the hole
are at distances $\ell$ apart.

\noindent
--take up a unit space in quasimomentum. Indeed, incrementing all
quasimomenta above a given place in the sea by {\it two} units
creates two holes in that place, and so on; by locally reshuffling 
quasimomenta we can then separate these holes.

Note that holes are {\it not} antiparticles. Removing a 
particle for the sea creates a gap of $\ell$ spaces and, from above,
$\ell$ holes. The correspondence is
\be
1~{\rm particle} \sim -\ell~{\rm holes}
\ee
We already observe a sort of duality between the two types of 
excitations. This can be summarized as
\be
{\rm particle} \leftrightarrow {\rm hole}~,~~~
\ell \leftrightarrow \frac{1}{\ell} ~,~~~
p \leftrightarrow \ell p
\ee
Under the above, the spectrum of excitations of the model remains
invariant. This is the simplest manifestation of a coupling-constant
duality that goes over to the correlation functions and Green's functions
of the model \cite{Gau,MPCF,LPS,Ha}. 
Obviously, this duality is spoiled by nonperturbative
effects, since holes are confined within the sea while there is no
`ceiling' for particle excitations.

This concludes our brief `physicist's tour' of the CSM model.
We are now going to examine the most useful ways of analyzing
this system.

\section{One-dimensional systems: Matrix model}

Identical particles can be formulated in terms of their
phase space variables modulo permutations. This leads to the
symmetric group approach of section 1. We could try other
approaches. For instance, we could formulate them in terms of
the eigenvalues of an $N \times N$ matrix. There is no a priori
ordering of these eigenvalues, so this certainly encodes identical
particles. It is clear that the permutation symmetry of the
problem has been promoted to the continuous symmetry of unitary
conjugations of this matrix, which leaves the eigenvalues intact.
This should be viewed as a gauge symmetry and would open the road to
yet another definition of statistics.

\subsection{Hermitian matrix model}

With this introductory remark as motivation, let us examine a
matrix model that parallels as closely as possible particle mechanics.
The kinematical variable is a hermitian $N \times N$ matrix $M$ and the
lagrangian reads 
\be
\LL = \tr \left\{ \half {\dot M}^2 - V(M) \right\}
\ee
$V(x)$ is a scalar potential evaluated for the matrix variable $M$.

Clearly the above has a time-translation invariance which leads to the
conserved energy
\be
H = \tr \left\{ \half {\dot M}^2 + V(M) \right\}
\ee
Moreover, the action is invariant under time-independent unitary 
conjugations of the matrix $M$:
\be
M \to U M U^{-1}
\ee
This nonabelian $SU(N)$ symmetry leads to the conserved  hermitian
traceless matrix
\be
J = i [M, {\dot M} ]
\ee
where $[ ~,~ ]$ denotes ordinary matrix commutator. These are the
`gauge charges' that, when fixed, will determine the `statistics'
of the model. But let us first examine the system classically.
We are interested in the dynamics of the eigenvalue of $M$, so we
parametrize it as
\be
M = U \Lambda U^{-1}
\ee
where $U(t)$ is the unitary `angular' part of the matrix and $\Lambda (t)
= diag\{x_1 , \dots x_N \}$ are the eigenvalues. Clearly the conserved
quantity $J$ has to do with invariance under `rotations' of the angular
part of $M$ and thus corresponds to the `angular momentum' of $U(t)$.
We define the `gauge potential'
\be
A = - U^{-1} {\dot U}
\ee
$\dot M$, $J$ and the lagrangian $\LL$ become, in this parametrization,
\begin{eqnarray}
{\dot M} &=& U \left( {\dot \Lambda} + [ \Lambda , A ] \right) U^{-1}
\label{dotM} \\
J &=& i U \left( \left[ \Lambda, [ \Lambda , A ]
\right] \right) U^{-1} \equiv U K U^{-1} \\
\LL &=& \tr \left\{ \half {\dot \Lambda}^2 + [ \Lambda , A ]^2
- V( \Lambda ) \right\} \\
&=& \half \sum_{i=1}^N {\dot x}_i - \half \sum_{i,j=1}^N ( x_i - x_j )^2
A_{ij} A_{ji}
\label{LLU}
\end{eqnarray}
The matrix elements of $A$ and $K$ are related
\be
K_{jk} = i \left[ \Lambda, [ \Lambda , A ] \right]_{jk}
=  i ( x_j - x_k )^2 A_{jk}
\label{Kjk}
\ee
Finally, solving (\ref{Kjk}) for $A_{jk}$ and putting into (\ref{LLU}) 
we obtain
\be
\LL = \sum_i \half {\dot x}_i^2 + \half \sum_{i \neq j} \frac{
K_{ij} K_{ji} }{( x_i - x_j )^2} -\sum_i V( x_i )
\ee
The first two terms are kinetic, coming from ${\dot M}^2$, while the
last one is potential. Therefore, the hamiltonian $H$ is
\be
H = \sum_i \half p_i^2 + \half \sum_{i \neq j} \frac{
K_{ij} K_{ji} }{( x_i - x_j )^2} +\sum_i V( x_i )
\label{HxK}
\ee
Note that the eigenvalues are kinematically coupled by an inverse-square
type potential with the angular momentum degrees of freedom. The connection
of the matrix model to the Calogero model along the lines presented
here and below was first established in \cite{KKS}. Also, the hamiltonian
(\ref{HxK}) has been proposed independently of the matrix model
as an $SU(N)$-generalization of the classical Calogero system
\cite{GH,Woj}.

We can now examine special cases:

a) The most `gauge invariant' sector is,
of course, the one in which the angular momentum charges vanish, that is,
$J=0$. In that case, (\ref{HxK}) for $K=0$ becomes the hamiltonian of
non-interacting particles in an external potential $V(x)$. This would
be the case of `standard' particles.

b) For the next simplest case the angular momentum $J$ should be
as trivial as possible without vanishing. Only its eigenvalues are
really relevant, since we can always perform a time-independent unitary
transformation $V$ which would shift $U \to VU$ and would rotate
$J \to V J V^{-1}$. The simplest choice would be to take the eigenvalues of 
$J$ to be equal. Unfortunately, this is not possible since the traceless
condition would make them vanish. The simplest possible choice is to take
all the eigenvalues equal to $\ell$ except one, which would cancel the
trace by being $(1-N) \ell$. This can be written in terms of an
arbitrary column $N$-vector $v$ as
\be
J = \ell ( v v^\dagger -1) ~,~~~ v^\dagger v = N
\ee
in which case $K$ becomes
\be
K = \ell ( u u^\dagger -1) ~,~~~ u = U^{-1} v
\ee
{}From (\ref{Kjk}) we see that $K_{ii} = 0$ (no sum on $i$) and thus
\be
u_i u_i^* =1  ~~~ {\rm (no~sum)}
\ee
So the coefficient of the inverse-square potential in (\ref{HxK})
becomes
\be
K_{ij} K_{ji} = \ell u_i u_j^* ~ \ell u_j u_i^* = \ell^2 ~~~(i \neq j)
\ee
Finally, (\ref{HxK}) becomes
\be
H = \sum_i \half p_i + \sum_{i < j} \frac{
\ell^2 }{( x_i - x_j )^2} +\sum_i V( x_i )
\label{Hxell}
\ee
This is the Calogero model! The potential strength $g=\ell^2$ is
related to the conserved charge $\ell$. Quantum mechanically,
picking this charge will amount to a choice of statistics. We also get, 
at this stage, an arbitrary external potential $V(x)$.

c) More general choices of $J$ amount to more variety in its eigenvalues.
$K_{ij} K_{ji}$ now, in general, becomes time-dependent and the dynamics
more complicated. We postpone the discussion for the quantum case where
it will be shown that this corresponds to Calogero particles having also
internal degrees of freedom. This will be a generalization of the
discussion of the first section, with irreps of $SU(N)$ substituting
the irreps of $S_N$.

Now that we have this new approach we can use matrix technology to
demonstrate the integrability of the Calogero model \cite{KKS,OP}. 
For $V(x)=0$ the matrix motion becomes free and $\dot M$ is conserved.
The conjugation-invariant quantities
\be
I_n = \tr {\dot M}^n
\ee
are also conserved and in involution (the matrix elements of $\dot M$
are momenta and have vanishing Poisson brackets). From (\ref{dotM})
and (\ref{Kjk}) we have
\begin{eqnarray}
{\dot M}_{jk} &=& U^{-1} \left( \delta_{jk} {\dot x}_j - 
(1-\delta_{jk} ) \frac{ i K_{jk}}{ x_j - x_k } \right) U \\
&=& 
U^{-1} \left( \delta_{jk} \, {\dot x}_j - 
(1-\delta_{jk} ) \frac{ i u_j u_k^* }{ x_j - x_k } \right) U 
\end{eqnarray}
When the above expression is inserted in the trace $I_n = \tr {\dot M}^n$
clearly $U$ drops and products of the form $u_i u_j^* \, u_j u_k^*
\dots u_i^*$ will appear which reduce to powers of $\ell$. Therefore,
the $I_n$ reduce to expressions involving only $x_i$, ${\dot x}_i$
and the coupling constant $\ell$. These are the conserved integrals of
the Calogero model.

Starting from the matrix model the actual motion of the Calogero
model can be obtained. The solution for $M$ is
\be
M = B+Ct
\ee
for arbitrary matrices $B,C$. The conserved charge becomes
\be
J = i [M, {\dot M} ] = i [B,C] = i \ell (u u^\dagger -1)
\label{JBC}
\ee
By unitary transformations we can choose the phases of $u$ such
that $u_i =1$; choices for $B,C$, then, that satisfy (\ref{JBC}) are
\be
B_{jk} =  \delta_{jk} \, q_j ~,~~~
C_{jk} = \delta_{jk} p_j - (1-\delta_{jk} ) \frac{ i \ell }
{ q_j - q_k } 
\ee
$q_i$ and $p_i$ are the initial conditions for $x_i$ and
${\dot x}_i$ at time $t=0$. Diagonalizing, then, $M=B+Ct$ for the
above $B,C$ produces the motion of the system. Another choice is
\be
B_{jk} = \delta_{jk} a_j + (1-\delta_{jk} ) \frac{ i \ell }
{ P_j - P_k } ~,~~~ C_{jk} = \delta_{jk} P_j
\ee
$P_i$ and $a_i$ are asymptotic momenta and impact parameters.
For $t \to \pm \infty$ the off-diagonal elements of $B$ produce a 
perturbation of order $t^{-1}$ to the eigenvalues, so the motion is 
determined by the diagonal elements $a_i + P_i t$ alone. We recover the
result that the motion at asymptotic regions is the same as if the
particles were free.

We conclude by mentioning that the matrix model is also integrable and
solvable in the presence of a harmonic oscillator potential
$V(x) = \half \omega^2 x^2$. The non-hermitian matrix $Q={\dot M}
+ i\omega M$ evolves as
\be
Q(t) = e^{i\omega t} Q(0)
\ee
and the matrix $Q^\dagger Q$ is conserved. We leave it as an exercise
to derive the conserved integrals and the motion of the corresponding
Calogero problem.

External potentials with up to quartic dependence on $x$ also lead to
integrable, although not so solvable, models \cite{PolB}. 
It is an open question
to prove this is all there is, or to find yet more integrable potentials.
For the purposes of statistics this is a rather academic issue.

Finally, we may wonder what restricts us to one dimension. We chose
a model with one matrix, and its eigenvalues corresponded to coordinates
of particles on the line. We could, indeed, start with an appropriate model
with many matrices, which would reproduce particle motion in higher
dimensions \cite{PolC}. The integrability and solvability properties of 
such extended models, however, are much less pleasant. The question of
whether they represent a workable extension of statistics remains open.

\subsection{The unitary matrix model}

The hermitian matrix model works well for particles on the line
but has trouble representing particles on periodic spaces. The most
natural candidate for such models would be a unitary $N \times N$ matrix
$U$. Its eigenvalues are phases and naturally live on the circle. We start,
therefore, with a lagrangian that represents the invariant kinetic energy
on the space of such matrices:
\be
\LL = -\half \tr ( U^{-1} {\dot U} )^2
\ee
A potential could in principle be included but we are interested in the
translationally invariant case and will omit it. The treatment is similar
as before, and we just summarize the relevant facts. 

The lagrangian is,
in fact, invariant under separate left- and right-multiplications of $U$ 
by time-independent unitary matrices and and there are two
corresponding conserved matrix angular momenta $L$ and $R$:
\begin{eqnarray}
U \to  VU : ~~~~~~~L &=& i {\dot U} U^{-1} \\
U \to  UW^{-1} :~~~  R &=& -i U^{-1} {\dot U}
\end{eqnarray}
The unitary conjugation that preserves the eigenvalues corresponds to
$W=V$ and its generator is
\be
J = L + R = i [ {\dot U} , U^{-1} ]
\ee
The rest of the discussion is as previously. Parametrizing 
\be
U = V \Lambda V^{-1} ~~{\rm with}~~
\Lambda = diag \{ e^{i x_i} , \dots e^{i x_N} \}
\label{UVparam}
\ee
the hamiltonian becomes, after a few steps,
\be
H = \sum_i \half p_i^2 + \half \sum_{i \neq j} \frac{
K_{ij} K_{ji}}{4 \sin^2 \frac{x_i - x_j}{2} }
\ee
where, as before,
\be
K = V^{-1}  J V
\ee
Choosing $J=K=0$ reproduces free particles on the circle, while
choosing $J = \ell (u u^\dagger -1)$ we obtain $K_{ij} K_{ji} = \ell^2$
and we recover the Sutherland inverse-sine-square model
\be
H = \sum_i \half {\dot x}_i^2 + \half \sum_{i \neq j} \frac{
\ell^2}{4 \sin^2 \frac{x_i - x_j}{2} }
\ee

This model is integrable and solvable by the same techniques as
the hermitian one. The conserved invariant quantities are
\be
I_n = \tr L^n = \tr (-R)^n = \tr (i U^{-1} {\dot U})^n
\ee
and the solution is
\be
U = B e^{iCt}
\ee
with $B$ a unitary and $C$ a hermitian matrix satisfying
\be
BCB^{-1} - C = J
\ee
For the Sutherland case with $J=\ell (u u^\dagger -1)$, $u_i =1$,
$B,C$ become
\be
B_{jk} = \delta_{jk} e^{i q_j} ~,~~~
C_{jk} = \delta_{jk} \, p_j + (1-\delta_{jk} ) \frac{i\ell}
{e^{i(q_j - q_k )} -1}
\ee
where, clearly, $q_i$ and $p_i$ are initial positions and momenta.

We conclude by mentioning that, upon scaling $x \to \alpha x$,
$t \to \alpha^2 t$, the Sutherland model goes over to the free Calogero model. 
This is the `infinite volume' limit.

\subsection{Quantization and spectrum}

We will, now, perform the quantization of this system on a periodic
space, using the unitary matrix model. We begin by defining a canonical
momentum matrix conjugate to the `coordinate' $U$
\be
\Pi = \frac{\partial \LL}{\partial {\dot U}} = - U^{-1} {\dot U} U^{-1}
\ee
The Poisson brackets are
\be
\{ U_{jk} , \Pi_{lm} \} = \delta_{jm} \delta_{lk}
\label{UPi}
\ee
$\Pi$ is somewhat unpleasant, being neither unitary nor hermitian.
We prefer to work in terms of the hermitian matrices $L$ and $R$ defined
previously
\be
L = i {\dot U} U^{-1} = -iU\Pi ~,~~~
R = -i U^{-1} {\dot U} = i \Pi U
\ee
Using (\ref{UPi}) we derive the following Poisson brackets:
\begin{eqnarray}
\{ L_{jk} , L_{lm} \} &=& i ( L_{jm} \delta_{lk} - \delta_{jm} L_{lk} ) \\
\{ L_{jk} , R_{lm} \} &=& 0 \\
\{ R_{jk} , R_{lm} \} &=& i ( R_{jm} \delta_{lk} - \delta_{jm} R_{lk} ) 
\end{eqnarray}
The above is nothing but two copies of the $U(N)$ algebra in disguise.
To see this, expand the matrices $L$ and $R$ in the basis of the 
fundamental generators of $SU(N)$ $T^a$ plus the unit matrix:
\be
L = L^o + 2 \sum_{a=1}^{N^2 -1} L^a T^a 
\label{Loa}
\ee
\be
R = R^o + 2 \sum_{a=1}^{N^2 -1} R^a T^a
\label{Roa}
\ee
with $L^o$, $L^a$, $R^o$, $R^a$ numbers. Then use the $SU(N)$
commutation relations
\be
[ T^a , T^b ] = i f^{abc} T^c
\ee
as well as the normalization 
\be
\tr ( T^a T^b ) = \half \delta_{ab}
\ee
to show that the expansion coefficients satisfy the Poisson algebra
\begin{eqnarray}
\{ L^a , L^b \} &=&  f^{abc} L^c  \\
\{ L^a , R^b \} &=& 0 \\
\{ R^a , R^b \} &=&  f^{abc} R^c
\end{eqnarray}
while $L^o$, $R^o$ are central. Note that the $U(1)$ charges
\be
L^o = - R^o = \frac{1}{N} \tr (-i U^{-1} {\dot U}) = \frac{1}{N}
\sum_i {\dot x}_i
\ee
are essentially the total momentum of the system.

We are now ready to perform quantization. In the
$U$-representation, where states are functions of $U$, $\Pi$ becomes 
the matrix derivative $\Pi_{jk} = -i \delta_U$, acting as
\be
\delta_U \tr (UB) = B ~,~~~ \delta_U \tr (U^{-1} B) = - U^{-1} B U^{-1}
\ee
where $B$ is a constant matrix, and similarly on expressions containing
more $U$'s.
$L$ and $R$, upon proper ordering, are represented as
\be
L = -U \delta_U ~,~~~ R = \delta_U \, \cdot \, U
\ee
where in $R$ it is understood that we {\it first} act with the derivative
and {\it then} right-multiply the result by $U$. With this ordering,
$L$ and $R$ become the proper $U(N)$ operators acting as
\be
L \tr (UB) = -UB ~,~~~ L \tr (U^{-1} B) = B U^{-1} 
\ee
\be
R \tr (UB) = BU ~,~~~ R \tr (U^{-1} B) = - U^{-1} B 
\ee
It is also useful to express their action on arbitrary functions of $U$ as
\begin{eqnarray}
\tr (i\epsilon L) f(U) &= f((1-i\epsilon)U) - f(U) \\
\tr (i\epsilon R) f(U) &= f(U(1+i\epsilon)) - f(U) 
\end{eqnarray}
where $\epsilon$ is an arbitrary infinitesimal hermitian matrix,
emphasizing their role as generators of left- and right-multiplication
on $U$. Correspondingly, the operators $L^a$ and
$R^a$ satisfy the $SU(N)$ algebra. Their action can be obtained by taking
$\epsilon = \varepsilon T^a$ with $\varepsilon$ an infinitesimal 
scalar parameter, that is,
\begin{eqnarray}
i\varepsilon L^a f(U) &= f((1-i\varepsilon T^a) U) - f(U)
\label{elra} \\
i\varepsilon R^a f(U) &= f(U(1+i\varepsilon T^a)) - f(U) 
\end{eqnarray}
The hamiltonian, being classically
the kinetic term on the manifold of unitary matrices $U(N)$, quantum 
mechanically becomes the laplacian operator on the manifold \cite{NAM}.
Using (\ref{Loa},\ref{Roa}) it is expressed as
\be
H = \half \tr L^2 = \sum_a (L^a)^2 + \half N (L^o)^2 =
\sum_a (R^a)^2 + \half (R^o)^2 = \half \tr R^2 
\ee
It is, therefore, the common quadratic Casimir of the left- and
right-$SU(N)$ algebra plus the square of the $U(1)$ charge, the two
parts identifiable as the relative and center-of-mass energy respectively.

Quantum mechanical states grouping into irreducible representations
of the $L$ and $R$ $SU(N)$ algebras will, thus, be degenerate
multiplets of the hamiltonian. The $U(1)$ (center of mass) part 
trivially separates: we can boost any state by any desired total
momentum $NP$ by multiplying the wavefunction by $(\det U)^P$. We will
examine only the $SU(N)$ part from now on.

A natural basis of states for the Hilbert space are the matrix elements
of the unitary irreducible representations (irreps) of $SU(N)$. Let $R$ 
denote such an irrep, $R(U)$ the matrix that represents $U$ in this irrep
and $R_{\alpha \beta} (U)$ the $\alpha \beta$ matrix element of this matrix.
Clearly $\alpha$ and $\beta$ range from 1 to the dimensionality of
$R$, $d_R$. $R_{\alpha \beta} (U)$ are a complete orthonormal basis of
wavefunctions for $U$, that is
\be
\int [dU] R_{\alpha \beta} (U) R'_{\gamma \delta} (U)^* = \delta_{R R'}
\delta_{\alpha \gamma} \delta_{\beta \delta}
\ee
with $[dU]$ the volume element on the space of $SU(N)$ matrices as
implied by the metric $ds^2 = -\tr ( U^{-1} dU )^2$, also called the
Haar measure. 

We will, now, show that each $R_{\alpha \beta} (U)$ is an eigenstate
of the hamiltonian with eigenvalue equal to the quadratic Casimir of
$R$, $C_R$. Qualitatively, after the discussion of the last paragraphs,
this should be obvious: $L$ and $R$ generate the transformations
$U \to V^{-1} U$ and $U \to UW$. $R (U)$ transforms in the conjugate
irrep ${\bar R}$ under $L$ and in the irrep $R$ under $R$. Since $H$ 
is the common quadratic Casimir of $L$ and $R$ we conclude that all
$d_R^2$ states $R_{\alpha \beta} (U)$ are energy eigenstates with 
eigenvalue $C_R = C_{\bar R}$.

(If you are confused about $L$ generating 
$U \to V^{-1} U$ rather than $U \to VU$, think of
the difference between active and passive transformations, which is
relevant when shifting from classical to quantum: $\psi (x-a)$
shifts the wavefunction by $+a$. Also, although classical transformations
on $U$ compose properly,
\be
V_1 ( V_2 U) = (V_1 V_2 ) U
\ee
quantum mechanically the operators $\hat V$ that perform the shift
$U \to VU$ on the argument of the wavefunction would compose
\be
{\hat V}_1 ( {\hat V}_2 f(U)) = {\hat V}_1 f( V_2 U) 
= f( V_2 V_1 U) = ({\hat V}_2 {\hat V}_1) f(U)
\ee
Therefore we need to invert the action of $\hat V$ to get the right
composition law. See also the discussion of irreps of $S_N$ in the
section about permutation group statistics.)

Let us prove the fact $H R_{\alpha \beta} (U) = C_R R_{\alpha \beta} (U)$
more analytically. Since $R (U)$ is a representation, it obeys the
group property
\be
R_{\alpha \beta} (UV) = \sum_\gamma 
R_{\alpha \gamma} R_{\gamma \beta} (V)
\ee
{}From (\ref{elra}) we have
\begin{eqnarray}
(1+i\varepsilon ) L^a R_{\alpha \beta} (U) &=& R_{\alpha \beta} 
((1-i\varepsilon T^a) U) = R_{\alpha \gamma} (1-i\varepsilon T^a)
R_{\gamma \beta} (U) \\
&=& R_{\alpha \beta} (U) - i\varepsilon
R_{\alpha \gamma}^a R_{\gamma \beta} (U)
\end{eqnarray}
where $R^a = R(T^a )$ is the $a$-th generator of $SU(N)$ in the
$R$ representation. So
\be
L^a R_{\alpha \beta} (U) = -R_{\alpha \gamma}^a R_{\gamma \beta} (U)
\ee
and
\be
\sum_a (L^a )^2 R_{\alpha \beta} (U) = \sum_a R_{\alpha \gamma}^a 
R_{\gamma \delta}^a R_{\delta \beta} (U) = 
\sum_a (R^a )_{\alpha \delta}^2 R_{\delta \beta} (U) 
\ee
The sum $\sum_a (R^a )^2$ appearing above is the quadratic
Casimir in the irrep $R$ and is proportional to the identity matrix
$\delta_{\alpha \delta}$. So, finally,
\be
H R_{\alpha \beta} (U) = C_R R_{\alpha \beta} (U)
\ee
Incidentally, the spectrum spanned by $C_R$ for all $R$ is nothing but
the spectrum of $N$ free fermions on the circle with the ground
state energy and the center-of-mass energy subtracted, where the 
lengths $R_i$ of the rows of the Young tableau of $R$ correspond 
to the ``bosonized'' fermion momenta
\be
p_i = R_i -i+1
\ee
and where the center-of-mass energy has been subtracted. The condition
$R_i \ge R_{i+1}$ for the rows amounts to the fermionic condition
$p_i > p_{i+1}$. The spectrum of the full matrix model, then, 
is identical to the free fermion one but with
{\it different degeneracies}.

We have, therefore, identified all energy eigenstates of the matrix model.
It remains to implement the quantum analog of the choice of angular
momentum $J$, identify the corresponding reduced quantum model, and pick
the subspace of states of the full model that belongs to the reduced
model.

$J$ obeys itself the $SU(N)$ algebra (it is traceless, no $U(1)$ charge).
A choice of value for $J$ amounts to a choice of irrep $r$ for this algebra.
States within the same irrep are related by unitary transformations
of $U$ and give the same dynamics; they will be discarded as gauge
copies, and only the choice of irrep will be relevant. Since $J=L+R$,
we see that states transforming under $(L,R)$ in the $({\bar R}, R)$
irreps will transform in the ${\bar R} \times R$ under $J$. So, only
irreps $r$ that are contained in the direct product of two mutually
conjugate irreps can be obtained for $J$. This amounts to irreps $r$
with a number of boxes in their Young tableau that is an integer multiple
of $N$. (To get a feeling of this, consider the case $N=2$. Then
$J$ is an orbital-like realization of the angular momentum through
derivatives of $U$ and clearly cannot admit spinor representations.)

We must, therefore, project the $d_R^2$ states in $R_{\alpha \beta} (U)$
to the subspace of states transforming as $r$ under $L+R$.
Call $G({\bar R}, \alpha ; R, \beta | r , \gamma )$ the Clebsch-Gordan 
coefficient that projects these states to the $\gamma$ state of $r$.
Then the relevant states for this model become
\be
\Psi_R ( U ) = \sum_{\alpha , \beta} R_{\alpha \beta} (U) G({\bar R}, \alpha;
R, \beta | r , \gamma ) 
\ee
The index $\gamma$ labeling the states within $r$, as we argued before,
counts the $d_r$ gauge copies and does not imply a true degeneracy of states.
The degeneracy of the states produced by each $R$ is, then, given by the
number of times that the irrep $r$ is contained in the direct
product ${\bar R} \times R$ or, equivalently, the
number of times that $R$ is contained in $R \times r$. Calling this 
integer $D( R,r ; R )$, we obtain for the spectrum and degeneracies:
\be
E_R = C_R ~,~~~ D_R = D( R,r ; R )
\label{spectrum}
\ee
In particular, if $D_R =0$ the corresponding energy level is
absent from the spectrum.

Concluding, we mention that an approach which also reproduces the
spectrum and states of the Sutherland model is two-dimensional
Yang-Mills theory on the circle \cite{GN,MPYM}. This approach
is essentially equivalent to the matrix model above and we
will not be concerned with it.

\subsection{Reduction to spin-particle systems}

So we have derived the spectrum, degeneracy and wavefunctions of the
matrix model restricted to the sector $J=r$. Classically these
restrictions represented free particles ($J=0$), Sutherland particles
($J= \ell (v v^\dagger -1)$) or something more general. What are the
corresponding quantum systems?

To find these, let us reproduce here the expression of the reduced
hamiltonian in one of these sectors:
\be
H = \sum_i \half p_i^2 + \half \sum_{i \neq j} \frac{
K_{ij} K_{ji}}{4 \sin^2 \frac{x_i - x_j}{2} } -E_o
\label{Hquant}
\ee
This expression remains valid quantum mechanically upon a proper
definition (ordering) of the operator $K$. The only residual quantum
effect is a constant term $E_o$ that comes from the change of measure
from the matrix space to the space of eigenvalues.

Let us expand a bit on this without entering too deeply into the 
calculations. (For details se, e.g., \cite{Meh}.) The Haar measure 
in terms of the diagonal and angular part of $U$ has the form
\be
[dU] = \Delta^2 \, [dV]
\ee
where $[dV]$ is the Haar measure of $V$ and $\Delta$ is the 
Vandermonde determinant
\be
\Delta = \prod_{i<j} 2 \sin \frac{x_i - x_j }{2}
\ee
To see this, write the `line element' $-\tr (U^{-1} dU)^2$ in terms of
$V$ and $x_i$ using (\ref{UVparam}) and obtain
\be
-\tr (U^{-1} dU)^2 = \sum_i dx_i^2 - \sum_{i,j}
4 \sin^2 \frac{x_i - x_j }{2} (V^{-1} dV)_{ij} (V^{-1} dV)_{ji} 
\ee
This metric is diagonal in $dx_i$ and $(V^{-1} dV)_{ij}$. The 
square root of the determinant
of this metric, which gives the measure (volume element) on the space,
is clearly $\Delta^2$ times the part coming from $V$ which is the
standard Haar measure for $V$. (We get {\it two} powers of $4 \sin^2
\frac{x_i - x_j}{2}$ in the determinant, one from the real and one
from the imaginary part of $(V^{-1} dV)_{ij}$, so the square root
of the determinant has one power of $\Delta^2$.)

To bring the kinetic $x_i$-part into a `flat' form (plain second
derivatives in $x_i$) we must multiply the wavefunction with the
square root of the relevant measure (compare with the change from
cartesian to radial coordinate in spherical potential problems).
The net result is that the wavefunction $\Psi$ in terms of $x_i$ and
$V$ is the original wavefunction $\psi (U)$ of the matrix model
times the Vandermonde determinant. This, however, also produces
an additive constant $E_o$ which comes from the action of the entire
$x_i$-kinetic operator on $\Delta$. Noticing that $\Delta$ is 
nothing but the ground state wavefunction of $N$ free fermions on
the circle, we see that $E_o$ is the relevant fermionic ground 
state energy
\be
E_o = \frac{N(N^2 -1)}{24}
\ee
This is the famous `fermionization' of the eigenvalues produced by
the matrix model measure.

To determine the proper ordering for $K$ we examine its properties
as a generator of transformations. Since $U = V \Lambda V^{-1}$, 
and $J$ generates $U \to V' U V'^{-1} = (V' V) \Lambda (V' V)^{-1}$,
we see that $J$ generates left-multiplications of the angular part
$V$ of $U$. $K = V^{-1} J V$, on the other hand, generates {\it
right}-multiplications of $V$, as can be see from its form or
by explicit calculation through its Poisson brackets. As a result,
it also obeys the $SU(N)$ algebra. Its proper quantum definition, then,
is such that it satisfies, as an operator, the $SU(N)$ algebra.
It clearly commutes with the diagonal part $x_i$ and its momentum $p_i$,
since it has no action on it. Its dynamics are fully determined
by the hamiltonian (\ref{Hquant}) and its $SU(N)$ commutation relations.

We can, therefore, in the context of the particle model (\ref{Hquant}),
forget where $K$ came from and consider it as an independent set of 
dynamical $SU(N)$ operators. $K$, however, obeys some constraints.
The first is that, as is obvious from $K = V^{-1} J V$, $K$ carries 
{\it the same irrep $r$} as $J$. The second is subtler: a 
right-multiplication of $V$ with a diagonal matrix will clearly leave
$U = V \Lambda V^{-1}$ invariant. Therefore, this change of
$V$ has no counterpart on the `physical' degrees of freedom of the
model and is a gauge transformation. As a result, we get the `Gauss' law'
that physical states should remain invariant under such transformations.
Since $K$ generates right-multiplications of $V$, and $K_{ii}$ (no sum)
generates the diagonal ones, we finally obtain
\be
({\rm no~sum})~~K_{ii} = 0 ~~~({\rm on~physical~states})
\label{Kcon}
\ee
(A more pedestrian but less illuminating way to see it is: $J = i
[ U^{-1} , {\dot U} ]$, being a commutator, vanishes when sandwiched 
between the same eigenstate of $U$. Since $K$ is essentially $J$ in
the basis where $U$ is diagonal, its diagonal elements vanish.)
Note that the constraint (\ref{Kcon}) is preserved by the hamiltonian
(\ref{Hquant}).

The above fully fixes the reduced model Hilbert space as the product of
the $N$-particle Hilbert space times the $d_r$-dimensional space of $K$,
with the constraint (\ref{Kcon}) also imposed. The further casting of the
model into something with a more direct physical interpretation relies
upon a convenient realization of $K$. {\it Any} such realization will
do: simply break the representation of $SU(N)$ that it carries into
irreps $r$ and read off the spectrum for each $r$ from the results of
the previous section.

We shall implement $K$ in a construction {\it \`a la} Jordan-Wigner. Let 
$a_{mi}$, $a_{mi}^\dagger$, $m=1, \dots q$, $i=1, \dots N$ be a set of 
$Nq$ independent bosonic oscillators \cite{MPYM}:
\be
[ a_{mi} , a_{nj}^\dagger ] = \delta_{mn} \delta_{ij}
\ee
Then
\be
K^a = \sum_{m=1}^q a_{mi} T_{ij}^a a_{mj}
\ee
is a realization of the $SU(N)$ algebra. ($T_{ij}^a$ are the matrix
elements of $T^a$.) The corresponding matrix elements of $K$ are
\be
K_{ij} = \sum_{m=1}^q \left\{ a_{mi}^\dagger a_{mj} - \frac{1}{N} 
\left( \sum_k a_{mk}^\dagger a_{mk} \right) \delta_{ij} \right\}
\label{Kij}
\ee
Correspondingly, the coefficient of the Sutherland potential in 
(\ref{Hquant}) is (for $i \neq j$)
\be
K_{ij} K_{ji} = \sum_{m,n} a_{mi}^\dagger a_{ni} \, a_{nj}^\dagger 
a_{mj} + \sum_m a_{mi}^\dagger a_{mi}
\ee
We already see that the degrees of freedom of $K$ are redistributed into 
degrees of freedom for each particle in the above. Specifically, defining
\be
S_{i,mn} = a_{mi}^\dagger a_{ni} - \frac{1}{q} \left( \sum_{s=1}^q 
a_{si}^\dagger a_{si} \right) \delta_{mn}
\label{Smn}
\ee
and comparing with (\ref{Kij}) we see that the $S_i$ are $N$ independent
sets of operators each satisfying the $SU(q)$ algebra. Before expressing
$K_{ij} K_{ji}$ in terms of the $S_i$ let us see what the constraint
(\ref{Kcon}) implies:
\be
K_{ii} = \sum_{m=1}^q a_{mi}^\dagger a_{mi} - 
\frac{1}{N} \sum_{m,k} a_{mk}^\dagger a_{mk} = 0
\ee
$\sum_{m,k} a_{mk}^\dagger a_{mk}$ commutes with all $K_{ij}$
and all $S_{i,mn}$. It is, therefore, a Casimir and can be chosen as a 
fixed integer $\ell N$ equal to the total number operator of the subspace 
of the oscillator Fock space in which the model lives. The above
constraint, then, implies
\be
\sum_{m=1}^q a_{mi}^\dagger a_{mi} = \ell 
\label{Acon}
\ee
(We see why we had to choose the total number operator to be a
multiple of $N$: the operator in (\ref{Acon}) above is also a
number operator and can have only integer eigenvalues.)
Using this in (\ref{Smn}) we can express 
\be
a_{mi}^\dagger a_{ni} = S_{i,mn} + \frac{\ell}{q} \delta_{mn}
\ee
and therefore
\be
K_{ij} K_{ji} = \sum_{mn} S_{i,mn} S_{j,nm} + \frac{\ell (\ell+q)}{q}
= {\vec S}_i \cdot {\vec S}_j + \frac{\ell (\ell+q)}{q}
\ee
where ${\vec S}_i \cdot {\vec S}_i = \tr (S_i S_j )$ is the
$SU(q)$-invariant scalar product of the two $SU(q)$ `vectors.'
We finally obtain the hamiltonian as \cite{MPYM}
\be
H = \sum_i \half p_i^2 + \half \sum_{i \neq j} \frac{
2{\vec S}_i \cdot {\vec S}_j + \frac{\ell (\ell +q)}{q}}
{4 \sin^2 \frac{x_i - x_j}{2} }
\label{HSquant}
\ee
So it is a Sutherland-like model but where the particles also
carry $SU(q)$ internal degrees of freedom (`spins') and the potential
contains a pairwise antiferromagnetic interaction between the spins.

It remains to specify the representation in which the $SU(q)$ spins
are and find the irreps contained in this realization of $K$, therefore
obtaining the spectrum. A realization of the form (\ref{Kij}) for
$q=1$ in terms of bosonic oscillators contains all {\it totally symmetric}
irreps of $S(N)$ (that is, the ones with a single row in their Young
tableau). (\ref{Kij}) is essentially the direct product of $q$ such
independent realizations, so it contains all direct products of $q$
totally symmetric irreps. This produces all irreps with up to $q$
rows in their Young tableau, perhaps more than once each. The constraint
(\ref{Acon}), however, implies that the total number of boxes in the
Young tableau of these irreps is $\ell N$. We recover once more the
constraint that we derived before based on the origin of $r$ as a
component of ${\bar R} \times R$.

Similarly, the realization (\ref{Smn}) of $S_i$ contains all the
totally symmetric irreps of $SU(q)$. (\ref{Acon}) implies that the number
of boxes of these irreps is equal to $\ell$, so the spins $S_i$ are each
in the $\ell$-fold symmetric irrep of $SU(q)$. Solving this model
amounts to decomposing the tensor product of these $N$ spins into 
irreducible components of $SU(q)$. Each such component corresponds to
a subspace of the Hilbert space with a fixed total spin $S$. This same 
irrep, interpreted as an irrep $r$ of $SU(N)$, will be the corresponding
irrep of $K$, and also of $J$, and thus will determine the spectrum of 
this sector through (\ref{spectrum}).

Let us elucidate the above by reproducing the two simplest cases: free 
particles and (spinless) Sutherland particles, comparing with the classical
treatment. 

a) Free particles correspond to $J=K=0$. So there is no spin and no potential
and we have non-interacting particles. From (\ref{spectrum}) we see that all
$D_R$ are one, and thus the spectrum is the free fermion one, as commented
before. The matrix model naturally quantizes free particles as fermions.

b) Spinless Sutherland particles correspond, classically, to $J = \ell
(v v^\dagger -1)$. So $J$ is rank one (ignoring the trace). Quantum
mechanically this corresponds to the irrep $r$ of $J$ having only one
row and therefore only one independent Casimir. Since $q$ in the realization
above corresponds to the number of rows, we must have $q=1$. Spins,
therefore, are absent.
The strength of the potential becomes $\ell (\ell +1)$ where $\ell N$ is
the number of boxes in the one row of $r$. By standard Young tableau
rules we see that the degeneracy $D_R$ is one if the row lengths of $R$
satisfy
\be
R_i \ge R_{i+1} + \ell
\ee
else it is zero. The spectrum of this model is, then, the same as the
spectrum of free particles but with the selection rule for their momenta
\be
p_i \ge p_{i+1} + \ell +1
\ee
We recover the `minimum distance' selection rule of the CSM model that
led to the interpretation as particles with generalized statistics! Only,
in this case, the statistics parameter $\ell +1$ is a positive integer.

We mention here that a Jordan-Wigner realization of $K$ in terms of
{\it fermionic} oscillators is also useful and leads to particles with spins
interacting via {\it ferromagnetic} Sutherland-type potentials. The
hamiltonian becomes \cite{MPYM}
\be
H = \sum_i \half p_i^2 - \half \sum_{i \neq j} \frac{
2{\vec S}_i \cdot {\vec S}_j + \frac{\ell(\ell -q)}{q}}
{4 \sin^2 \frac{x_i - x_j}{2} }
\label{HFquant}
\ee
where now the spins are in the $\ell$-fold {\it anti}symmetric irrep of
$SU(q)$. We will not elaborate further and leave the details as an exercise
to the reader.

\vskip 0.5cm
In conclusion, the matrix model has provided us with the following:

1. An augmentation of the permutation group into the $SU(N)$ group
and a corresponding possibility to define statistics through the irreps
of $SU(N)$.

2. A realization of generalized scalar statistics but with {\it quantized}
statistics parameter $\ell +1$ in terms of the CSM model.

3. A realization of generalized `non-abelian statistics' in terms of
particles with internal degrees of freedom interacting through a generalized
CSM-type potential.

4. A systematic way of solving the above models.

\noindent
What the matrix models has {\it not} provided is

1. A realization of generalized statistics for fractional statistics 
parameter.

2. A realization of spin-CSM systems with the spins in arbitrary 
(non-symmetric) representations.

3. A control of the coupling strength of the potential for the spin-CSM
models. (Note that the coefficient of ${\vec S_i} \cdot {\vec S_j}$ terms
is fixed to $\pm 2$ and also the constant therm $\ell (\ell +q)$ is
entirely fixed by the spin representation.)

There exist generalizations of the above models both at the classical
\cite{BL,PolR} and the quantum level \cite{PolM}. They all share,
however, the limitations spelled out above, especially (3).
These restrictions are important, in the quest of more general
statistics but also from the more practical point of view of solving
spin-chain models with spins not in the fundamental representation, as we
will shortly explain. For this reason, yet a different approach will be
pursued in the next section, namely, the operator approach.

\section{Operator approaches}

The matrix model connection provided us with a powerful tool
that not only allowed us to generalize statistics but also
led to the full quantum solution of a set of spin-generalized
CSM models.

As noted, however, in the conclusion of the preceding lecture,
the matrix model fixes the coefficient of the spin-interaction
and scalar interaction terms to $\pm 2$ and $\pm \ell (\ell \pm q)$
respectively. We cannot choose these coefficients at will.

We would like to have an approach that defeats this restriction
and leads to spin models with arbitrary coupling strengths. (This
is necessary to attack spin-chain systems through the 
infinite-coupling limit trick to be explained later.) Such an approach
should also be able to bypass the excursion to matrix models and 
deal more directly with these systems in an algebraic way.
This will be achieved with the exchange operator formalism \cite{PolEX}
(also known as the Dunkl operator formalism \cite{Dunk}).

\subsection{Exchange operator formalism}

Consider the operators $M_{ij}$ that permute the {\it coordinate} 
degrees of freedom of $N$ particles in one dimension which could,
in principle, also have internal degrees of freedom ($M$ for {\it
metathesis}, to avoid confusion with momenta $p_i$). They satisfy
the permutation algebra (symmetric group), in particular
\begin{eqnarray}
&M_{ij} = M_{ij}^{-1} = M_{ij}^\dagger = M_{ji} \\
&[ M_{ij} , M_{kl} ] = 0 ~~~{\rm if}~i,j,k,l~{\rm distinct} \\
&M_{ij} M_{jk} = M_{ik} M_{ij} ~~~{\rm if}~i,j,k~{\rm distinct}
\end{eqnarray}
Any operator $A_i$ on the phase space satisfying
\begin{eqnarray}
M_{ij} A_k &=& A_k M_{ij} ~~~{\rm if}~i,j,k~{\rm distinct}\\
M_{ij} A_i &=& A_j M_{ij}
\end{eqnarray}
will be called a {\it one-particle} operator (even though
it may involve the coordinates and momenta of many particles).

We construct the following one-particle operators \cite{PolEX}:
\be
\pi_i = p_i + \sum_{j \neq i} i W(x_i - x_j ) M_{ij} \equiv
p_i + \sum_{j \neq i} i W_{ij} M_{ij} 
\ee
We shall view the $\pi_i$ as generalized momenta. To ensure their
hermiticity the {\it prepotential} $W(x)$ should satisfy
\be
W(-x) = -W(x)^*
\ee
We shall construct the corresponding `free' hamiltonian from $\pi_i$
\be
H = \sum_i \half \pi_i^2
\ee
In terms of the original $p_i$ this hamiltonian will, in general
contain linear terms. To ensure that such terms are absent we must
further impose
\be
W(-x) = -W(x) = {\rm real}
\ee
With the above restriction the hamiltonian $H$ and commutation relations
of the $\pi_i$ become
\be
[ \pi_i , \pi_j ] = \sum_k W_{ijk} (M_{ijk} - M_{jik} )
\ee
\be
H = \sum_i \half p_i^2 + \sum_{i<j} \left( W_{ij}^2 + W_{ij}' M_{ij}
\right) + \sum_{i<j<k} W_{ijk} M_{ijk}
\ee
where we defined the three-body potential and cyclic permutation
\begin{eqnarray}
W_{ijk} &=& W_{ij} W_{jk} + W_{jk} W_{ki} + W_{ki} W_{ij} \\
M_{ijk} &=& M_{ij} M_{jk} 
\end{eqnarray}
To obtain an interesting and tractable model, $W_{ijk}$, which
appears in the commutator $[\pi_i , \pi_j ]$ and also as a three-body
potential, should vanish or at most be a constant. This leads to
a functional equation for $W(x)$:
\be
W(x) W(y) - W(x+y) \left[ W(x)+W(y) \right] = {\rm const} (= W_{ijk} )
\ee
We present the solutions:

a) $W_{ijk} = 0 ~\rightarrow~ W(x) = \ell /x$

b) $W_{ijk} =-\ell^2 <0 ~\rightarrow~ W(x) = \ell \cot x$

c) $W_{ijk} = +\ell^2 >0 ~\rightarrow~ W(x) = \ell \coth x$

\noindent
Let's examine each case.

a) In this case the $\pi_i$ become \cite{Dunk,PolEX,BHV}
\be
\pi_i = p_i + \sum_{j \neq i} \frac{i\ell}{x_{ij}} M_{ij} 
\ee
and satisfy
\be
[ \pi_i , \pi_j ] =0
\ee
The $\pi_i$ commute, so we can consider them as independent momenta.
The hamiltonian reads
\be
H = \sum_i \half p_i^2 + \sum_{i<j} \frac{\ell (\ell -M_{ij} )}
{x_{ij}^2}
\ee
We obtain a Calogero-like model with {\it exchange interactions}.
Yet it is nothing but a free model in the commuting momenta $\pi_i$.
Integrability is immediate: the permutation-invariant quantities
\be
I_n = \sum_i \pi_i^n
\ee
obviously commute with each other. If we assume that the particles
carry no internal degrees of freedom and are bosons or fermions then
$M_{ij} = \pm 1$ on physical states. The model becomes the standard
Calogero model and we have proved its integrability in one scoop.
(You may be left with a question mark: the hamiltonian and the
other integrals $I_n$ become the standard Calogero ones if $M_{ij}
= \pm 1$, so these reduced integrals will commute on the bosonic or 
fermionic subspace; but will they also commute on the {\it full}
Hilbert space? Prove for yourself that this is indeed the case.)

We can also construct harmonic oscillator operators \cite{PolEX,BHV}.
The commutators between $x_i$ and $\pi_i$ are
\begin{eqnarray}
{[}x_i , \pi_i {]} &=& i \left( 1+ \ell \sum_{j \neq i}  M_{ij} 
\right) \\
{[}x_i , \pi_j {]} &=& -i\ell M_{ij} ~~~ (i \neq j)
\end{eqnarray}
Defining
\begin{eqnarray}
a_i &=& \frac{1}{\sqrt 2} \left( \pi_i - i\omega x_i \right) \\
a_i^\dagger &=& \frac{1}{\sqrt 2} \left( \pi_i + i\omega x_i \right) 
\end{eqnarray}
we can show
\begin{eqnarray}
{[} a_i , a_i^\dagger {]} &=& \omega \Bigl( 1+\ell \sum_{j \neq i} 
M_{ij} \Bigr) \\
{[} a_i , a_j^\dagger {]} &=& -\omega \ell M_{ij} ~~~(i \neq j) \\
{[} a_i , a_j {]} &=& {[} a_i^\dagger , a_j^\dagger {]} = 0
\end{eqnarray}
This is an extended version of the Heisenberg algebra involving the
permutation operators. The corresponding oscillator hamiltonian reads
\be
H = \sum_i \half ( a_i^\dagger a_i + a_i a_i^\dagger ) =
\sum_i \half p_i^2 + \sum_i \half \omega^2 x_i + \sum_{i<j}
\frac{\ell (\ell -M_{ij} )}{x_{ij}^2}
\ee
and satisfies
\be
[ H , a_i ] = \omega a_i ~,~~~ [ H , a_i^\dagger ] =\omega a_i^\dagger
\label{Hacomm}
\ee
This is the harmonic Calogero model with exchange interactions,
which becomes again the standard model on bosonic or fermionic subspaces
for particles without internal degrees of freedom. Since
\be
H = \sum_i a_i^\dagger a_i + \half N \omega + \half \ell \omega 
\sum_{i \neq j} M_{ij}
\ee
we see that on bosonic or fermionic spaces the state annihilated by all
$a_i$ (if it exists) will be the ground state. Solving $a_i \psi =0$
we obtain for the ground state wavefunction
\begin{eqnarray}
\psi_B &=& \prod_{i<j} |x_{ij} |^\ell e^{-\half \omega \sum_i x_i^2} \\
\psi_F &=& \prod_{i<j} \left\{ sgn(x_{ij} ) |x_{ij} |^{-\ell} 
\right\} e^{-\half \omega \sum_i x_i^2} 
\label{BFgr}
\end{eqnarray}
For $\ell >0$ the bosonic state is acceptable, while for $\ell <0$
the fermionic one is acceptable. In the ``wrong'' combinations of
statistics and sign of $\ell$ the ground state is not annihilated
by the $a_i$, but it is still annihilated by all permutation-invariant
combinations of the $a_i$. 

{}From (\ref{Hacomm}) we see that we can find the spectrum of this model
for fermions or bosons by acting on the ground state with all
possible permutation-symmetric homogeneous polynomials in the 
$a_i^\dagger$. A basis bor these is, e.g.,
\be
A_n = \sum_i (a_i^\dagger )^n
\ee
So the spectrum is identical to non-interacting fermions or bosons,
but with a different ground state energy. For the `right' combinations
of $\ell$ and statistics, where (\ref{BFgr}) are the correct ground
state wavefunctions, the ground state energy is
\be
E_o = \frac{N}{2} \omega + \frac{N(N-1)}{2} |\ell | \omega
\ee
which is the correct Calogero result.

Finally, the quantities 
\be
I_n = \sum_i h_i^n = \sum_i (a_i^\dagger a_i )^n
\ee
can be shown to commute \cite{PolEX}, and therefore this system is 
also integrable.
It is left as an exercise to find the commutation relations of the $h_i$
and show that $[I_n ,I_m ] =0$.

b) In the case $W(x) = \ell \cot x$ we have 
\be
\pi_i = p_i + i \cot x_{ij} M_{ij}
\ee
\be
[ \pi_i , \pi_j ] = - \ell^2 \sum_k ( M_{ijk} - M_{jik} )
\ee
so the momenta are now coupled. The hamiltonian becomes
\be
H = \sum_i \half p_i^2 + \sum_{i<j} \frac{\ell (\ell -M_{ij} )}
{\sin^2 x_{ij}} - \ell^2 \left( \frac{N(N-1)}{2} + \sum_{i<j<k}
M_{ijk} \right)
\ee
We obtain the Sutherland model with exchange interactions plus
an extra term. On bosonic or fermionic states this becomes an
overall constant and we recover the standard Sutherland model.
Again, since $H$ is by construction positive definite, if a state
satisfying $\pi_i \psi =0$ exists it will be the ground state.
We obtain
\begin{eqnarray}
\psi_B &=& \prod_{i<j} |\sin x_{ij} |^\ell \\
\psi_F &=& \prod_{i<j} sgn(x_{ij}) |\sin x_{ij} |^\ell 
\end{eqnarray}
which are acceptable for the same combinations of $\ell$ and
statistics as before. For both cases $M_{ijk} =1$ so
\be
E_o = \ell^2 \frac{N(N^2 -1)}{24}
\ee
is the correct Sutherland model ground state energy. The excited
states can again be obtained in a (rather complicated) algebraic way
\cite{LV}. Finally, the quantities
\be
{\tilde \pi}_i = \pi_i + \ell \sum_{j \neq i} M_{ij} = p_i + 
e^{ix_i} \sum_{j \neq i} \frac{2\ell}{e^{ix_i} - e^{ix_j}} M_{ij}
\ee
can be shown to have the same commutation relations as the $h_i$
defined previously for the harmonic system. Therefore, the integrals
constructed from them
\be
I_n = \sum_i {\tilde \pi}_i^n
\ee
commute and the model is integrable.

c) For $W(x) = \ell \coth x$ we have a similar commutation
relation and a hamiltonian
\be
H = \sum_i \half p_i^2 + \sum_{i<j} \frac{\ell (\ell -M_{ij} )}
{\sinh^2 x_{ij}} + \ell^2 \left( \frac{N(N-1)}{2} + \sum_{i<j<k}
M_{ijk} \right)
\ee
This is the inverse-hyperbolic-sine-square model and supports
only scattering states. Its integrability can be obtained as
for the Sutherland model above, or simply as an `analytic 
continuation' of that model for imaginary period of space.
We will not examine it any further.

In conclusion, an exchange-family of models was introduced, solved and
related to the standard CSM models in spaces of definite symmetry.
It is remarkable that all these proofs work directly, and only, at
the quantum domain (there is no classical analog of $M_{ij}$).

\subsection{Systems with internal degrees of freedom}

We can easily extend the previous results for particles with
internal degrees of freedom. For this, assume that the particles are
{\it distinguishable} or, equivalently, that they carry a number $q$
of (discrete) internal degrees of freedom (species) that can be used to 
(partially) distinguish them. Their states are spanned by $|x,\sigma>$,
where $\sigma=1, \dots q$ counts internal states. The total permutation
operator $T_{ij}$, then is 
\be
T_{ij} = M_{ij} \sigma_{ij}
\ee
where $\sigma_{ij}$ is the operator that permutes the internal states
of particles $i$ and $j$.

Let us, then, simply take states that are bosonic or fermionic
under total particle exchange: $T_{ij} = \pm 1$. On such states
\be
M_{ij} = \pm \sigma_{ij}
\ee
and the Calogero and Sutherland exchange model hamiltonians become
\cite{MPSC}
\be
H_c = \sum_i \half p_i^2 + \sum_i \half \omega^2 x_i + \sum_{i<j}
\frac{\ell (\ell \mp \sigma_{ij} )}{x_{ij}^2} 
\ee
\be
H_s = \sum_i \half p_i^2 + \sum_{i<j} \frac{\ell (\ell \mp\sigma_{ij} )}
{\sin^2 x_{ij}} - \ell^2 \left( \frac{N(N-1)}{2} + \sum_{i<j<k}
\sigma_{ijk} \right)
\ee
We get the Calogero and Sutherland models with spin-exchange
interactions. From the completeness
relation for the fundamental $SU(q)$ generators $T^a$
\be
\sum_{a=1}^{q^2 -1} T_{\alpha \beta}^a T_{\gamma \delta}^a = 
\half \delta_{\alpha \delta} \delta_{\gamma \beta}
-\frac{1}{2q} \delta_{\alpha \beta} \delta_{\gamma \delta}
\ee
we deduce the form of the operators $\sigma_{ij}$
\be
\sigma_{ij} = 2 {\vec S}_i \cdot {\vec S}_j + \frac{1}{q}
\ee
where $S_i^a$ acts as $T^a$ on the internal states of particle $i$.
So the spin-dependent interaction coefficient of the potential 
in the hamiltonian takes the form \cite{HH,Kaw,MPSC,HW}
\be
\mp \ell \left( 2{\vec S}_i \cdot {\vec S}_j \mp \ell +
\frac{1}{q} \right)
\ee
We have recovered the ferromagnetic and antiferromagnetic spin
model of the previous section but with {\it arbitrary} coefficient!
On the other hand, the spins are necessarily in the fundamental
of $SU(q)$. So we have obtained a generalization of the coupling
constant with respect to the matrix model but a restriction of the
allowed spins. 

Note that $\ell$ here is an arbitrary parameter, while $\ell$ in
(\ref{HSquant}) was the size of the symmetric representation of $S_i$.
For $\ell=1$ and spins in the fundamental, the matrix model and 
exchange-operator model agree. It is interesting to note that we 
can go from ferromagnetic to antiferromagnetic interactions either 
by changing the sign of $\ell$ or by changing the statistics
of the particles.

The solution of the above models can be obtained algebraically.
For the spin-Sutherland model this is rather complicated and is
related to the so-called Yangian symmetry \cite{Hetc,BGHP,Cher,AJ}. 
For the spin-Calogero model it is easier \cite{PolD}. 
Let us concentrate on the model with 
interaction $\ell (-2{\vec S}_i \cdot {\vec S}_j \mp \ell -
\frac{1}{q} )$ and define the operators
\be
A_n^\dagger = \sum_i (a_i^\dagger )^n ~,~~~
(A_n^a )^\dagger = \sum_i (a_i^\dagger )^n S_i^a
\ee
and their hermitian conjugates.
These form a complete set for all permutation-symmetric creation
and annihilation operators for all species of particles. Yet the
commutators 
among themselves and with $H$ do not involve $\ell$. They create,
therefore, the same spectrum of excitations over the ground state
as $N$ noninteracting bosons or fermions with $q$ species.
For $\ell >0$ the ground state is the bosonic one:
\be
\psi_B = \prod_{i<j} |x_{ij} |^\ell e^{-\half \omega \sum_i x_i^2}
\chi_s (\{ \sigma_i \})
\ee
where $\chi_s$ is a totally symmetric state in the $\sigma_i$.
The set of all $\chi_s$ forms the $N$-fold symmetric irrep
of the total spin $S = \sum_i S_i$. Therefore the ground state 
is $(N+q-1)! / N! (q-1)!$ times degenerate. For $\ell<0$ the
above is not normalizable any more. But we remember that we can
obtain the same model by starting from fermions and the opposite 
coupling $-\ell >0$. The ground state, then, is of a fermionic
type
\be
\psi_F = \sum_P (-1)^P 
\left(\prod_i \delta_{\sigma_i ,\alpha_i} \right)
\prod_{i<j} |x_{ij} |^{-\ell} x_{ij}^{\delta_{\alpha_i ,\alpha_j}} 
e^{\half \omega \sum_i x_i^2}
\ee
where $P$ are total particle permutations and $\alpha_i$ are a
set of fixed values for the indices $\sigma_i$ that determine the
state. Clearly the ground state will be obtained for the minimal
total power of $x_i$ appearing above, and that will happen for
a maximally different set of values $\alpha_i$. These states form
the $n$-fold antisymmetric irrep of the total spin $S$, where $n
= N(mod q)$. The ground state is, thus, $q!/n!(q-n)!$ times degenerate.
The above spectra will come handy later.

\subsection{Asymptotic Bethe Ansatz approach}

We already mentioned that there are elaborate algebraic approaches
to derive the spectrum of the spin-Sutherland model, based on the
Yangian symmetry. We will, instead, take a lower-key approach which
reproduces the same spectra and is physically more lucid, although
not as rigorous. We will take the ABA route.

Consider {\it distinguishable} particles of the exchange-Calogero 
type without external potential, coming in with asymptotic
momenta $k_i$ and scattering off each other. Before scattering,
their positions are in some definite ordering determined by 
the ordering of their momenta (it is the inverse of that ordering).

The key observation is that, after scattering, the particles have
simply `gone through' each other with {\it no} backscattering \cite{SS}.
The impenetrable $1/x^2$ potential has become completely
penetrable in the presence of the exchange term! You can prove
this fact by examining the asymptotic properties of a simultaneous
eigenstate of $\pi_1 , \dots \pi_N$ which is obviously an energy
eigenstate: at $x_i \to \pm \infty$ the prepotential terms are vanishing
and we simply have eigenstates of the individual $p_i$. Since there are 
{\it no} pieces with the values of $p_i$ permuted (coming from
backscattering) we have complete transmission. 
(To explicitly see how it works, it is instructive to consider the 
two-body problem, decompose it into symmetric and antisymmetric parts,
scatter and recombine the parts after scattering. A relative
phase of $\pi$ between the two parts is what produces the effect.)

(Puzzle: what happens with the correspondence principle? 
With $\hbar$ back in, the interaction coefficient
is $\ell (\ell - \hbar M_{ij} )$. How can a term of order $\hbar$
produce such a dramatic effect, particles going through each
other, in the $\hbar \to 0$ limit?)

So the only effect of the scattering is a phase shift
of the wavefunction which, as we have said, is the sum of two-body
phases
\be
\theta_{sc} = \frac{N(N-1)}{2} \pi \ell
\ee
This is true on an infinite space. On a periodic space we can still
use the above result, together with the requirement for periodicity for
the wavefunction, to derive the spectrum. This is the ABA method and
is expected to reproduce the correct results in the thermodynamic limit
of many particles at constant density \cite{SS}.
It gives, in fact, the {\it exact} answer for the Sutherland model
\cite{Suth}, so we can expect it to work also in the present case.
For a space of period $2\pi$ the result is
\be
2\pi k_i + \sum_\pi \ell sgn(k_i - k_j ) = 2\pi n_i
\ee
The left hand side counts the total
phase picked up by a particle going round the space and scattering off 
to the other particles in the way. $n_i$ are arbitrary integers, ensuring
periodicity. There are, however, some constraints on the choice of
of $n_i$ that are imposed by continuity from the $\ell =0$ case:

\noindent
--If $k_i \le k_j$ then $n_i \le n_j$

\noindent
--If $n_i = n_j$ there is a {\it unique} solution, that is,
$k_i < k_j$ and $k_i > k_j$ represent the same state.

\noindent
These rules are important to avoid overcounting and to discard
spurious solutions. With these, the spectrum obtained is the same
as the one derived with more rigorous methods. For the ordering
$n_1 \le \dots n_N$ the solution for $k_i$ is
\be
k_i = n_i + \ell (i-\frac{N+1}{2} )
\label{ABAk}
\ee
and similarly for other orderings. We see that the ABA momenta $k_i$ are
the same as the quasimomenta that we have previously defined.

The bottom line is that the spectrum and degeneracies are the
same as those of distinguishable particles obeying generalized 
selection rules for their momentum. Still, what fixes the degeneracy
of states is the different ways that we can distribute the particles
to the quantum numbers $n_i$, rather than $k_i$ (see, especially,
the second rule above). A state of $N$ particles with the same
$n_i$, for instance, is nondegenerate although they, seemingly,
have different $k_i$ which would imply a permutation degeneracy.

For particles with spin the construction above, in combination with
the trick of the previous subsection of starting with fermions or
bosons, produces a spectrum with degeneracies the same as those
of free particles (the $n_i$ are `free' quantum numbers). As argued
before, for ferromagnetic interactions we must choose bosons and
combine their spins accordingly, while for antiferromagnetic 
interactions we must choose fermions. To spell it out, this means
the following:

\noindent
1. Choose a set of quantum numbers $n_i$. The ordering is immaterial,
since we have identical particles, so you can choose $n_1 \le \dots n_N$.

\noindent
2. Place your particles on these quantum numbers and put their
spins in the appropriate state. For the ferromagnetic case treat
them as bosons: the total spin of particles with the same $n_i$ transforms
in the symmetric tensor product of their spins. For the antiferromagnetic
case treat them as fermions: the total spin of of particles with the
same $n_i$ transforms in the antisymmetric tensor product of their
spins; clearly up to $q$ can have the same $n_i$ in this case.

\noindent
3. Calculate the energy of this state in terms of the ABA momenta
(\ref{ABAk}): $E = \sum_i k_i^2$.

\noindent
It should be obvious that similar rules applied to the spin-Calogero
system reproduce the spectrum derived in the last subsection.
This method can be used to calculate both the statistical mechanics
(large $N$) of these systems and the few-body spectra.

\subsection{The freezing trick and spin models}

Now that we have a tractable way of solving spin-CSM systems with
arbitrary strength of interaction we can introduce the freezing
trick \cite{PolE} and deal with spin chain models.

Consider, first, the previous ferromagnetic or antiferromagnetic
spin-Sutherland model. Take the limit $\ell \to \infty$.
The potential between the particles goes to infinity, so for
any finite-energy state the particles will be nearly `frozen' to 
their classical equilibrium positions. In fact, even the excitation
energies around that configuration will go to infinity:
the ground state energy scales like $\ell^2$, while the excitations
scale like $N \ell n + n^2$ with $n$ some excitation parameter.
So, to leading order in $\ell$ the spectrum becomes linear and of order 
$\ell$. These excitations correspond, essentially, to phonon modes
of small oscillations around the equilibrium positions of particles. The
`stiffness' of oscillations is, of course, proportional to the 
strength of the potential $\ell^2$ and the spectrum is proportional
to the frequency, of order $\ell$.

The quantum fluctuations of the particle positions in any state 
will scale like the inverse square root of the oscillator frequency,
that is, like $1/\sqrt \ell$. But, in the hamiltonian,
the piece coupling the spins to the kinematical degrees of freedom
is proportional to $1/ \sin ^2 x_{ij}$. In the large-$\ell$ limit, thus,
this term becomes a constant equal to its classical equilibrium value; 
so, in that limit, spin and kinematical degrees of freedom
decouple. (Note that the spin part is also of order $\ell$.)
The hamiltonian becomes
\be
H = H_S + \ell H_{spin}
\ee
with $H_S$ the spinless Sutherland hamiltonian and $H_{spin}$
the spin part
\be
H_{spin} = \mp \sum_{i<j} \frac{ 2 {\vec S}_i \cdot {\vec S}_j}
{4\sin^2 \frac{{\bar x}_{ij}}{2}}
\label{Hsp}
\ee
where the classical equilibrium positions ${\bar x}_j$ are equidistant
points on the circle:
\be
{\bar x}_j = \frac{2\pi j}{N}
\ee

The hamiltonian (\ref{Hsp}) above describes a spin chain consisting
of a regular
periodic lattice of spins in the fundamental of $SU(q)$ coupled
through mutual ferro- or antiferromagnetic interactions of strength
inversely proportional to their chord distance. It is the well known
$SU(q)$ Haldane-Shastry (HS) model \cite{Hal,Sha}. 
According to the above, its spectrum
can be found by taking the full spectrum of the corresponding
spin-Sutherland model in the large-$\ell$ limit, `modding out'
the spectrum of the spinless model and rescaling by a factor $1/\ell$.
Each state will inherit the spin representation of its `parent' 
spin-Sutherland state. So, both the energy and the total spin of
the states of the HS model can be determined this way. Commuting
integrals of this model \cite{Ino} can also be obtained this way
\cite{FM}. At the level of the partition function at some 
temperature $T$ we have
\be
Z_{spin} (T) = \lim_{\ell \to \infty} \frac{Z (\ell T)}
{Z_S (\ell T)}
\ee
{}From this, the thermodynamics of the spin chain model can be 
extracted \cite{SS}.

We will not give the details of this construction here. We urge
anyone interested to solve this way a few-site (two or three) spin
chain, see how it works and deduce the `construction rules' for the
spectrum of a general spin chain. Let us
simply state that the many degeneracies of the spectrum of the
HS model (larger that the total spin $SU(q)$ symmetry
would imply), which is algebraically explained by the existence of
the Yangian symmetry, can, in this approach, be explained in terms
of the degeneracies of free particles. (The degeneracies are not 
{\it identical}, due to the modding procedure, but related.)

For the spin-Calogero model a similar limit can be taken, scaling also
the external oscillator frequency as $\omega \to \ell \omega$ to keep
the system bound. The classical equilibrium positions of 
this model are
at the roots of the $N$-th Hermite polynomial. We obtain, therefore,
a non-regular lattice of spins interacting with a strength inversely
proportional to the square of their distance \cite{PolE}. The spectrum
of this model can be found quite easily with the above method. Again,
we refer to the literature for details \cite{PolD,Fra}.

In the continuum limit ($N \to \infty$) the antiferromagnetic version
of both the above models become $c=1$ conformal field theories, the HS
containing both chiral sectors while the inhomogeneous harmonic
one containing just one sector.

Other models exist and can be solved in this spirit: hierarchical
(many-coupling) models \cite{KwK}, supersymmetric models \cite{KtK},
`twisted' models \cite{FK} etc. 
All, however, work only for the fundamental representation of some
internal group. The big, important open problem is to crack a particle
system with a {\it higher} representation for the spins and 
{\it arbitrary} coupling strength. If this is done, through the
freezing trick we will be able to solve a spin chain with spins
in a higher representation. This is interesting since we could then
see if the antiferromagnetic system for integer $SU(2)$ spins develops
a mass gap, according to the Haldane conjecture \cite{HG}.

\section{Exclusion statistics}

So far we approached the problem of statistics in a
`fundamental' way, trying to give a reasonable definition
and presenting systems that realized this definition. In this
last section we shall give a `phenomenological' approach,
based on the state-counting properties of a system whose
dynamics may remain, otherwise, undetermined. It is
based on a notion already familiar from the CSM model, the
notion of `state repulsion' or `exclusion.' This will lead
to Haldane's `exclusion statistics.'

\subsection{Motivation from the CSM model}

Although Haldane derived his definition from the properties
of the HS spin chain, we will use the CSM model instead. 
Consider the `principal' quantum number of this model,
that is, the one in terms of which the energy eigenvalues
are a sum of independent terms: quasiexcitation numbers $n_i$
for the Calogero and quasimomenta $p_i$ for the Sutherland model.
They obey an `enhanced' exclusion principle where nearby values
can be no closer that $\ell$ units. It seems as if each of them
occupies $\ell$ places in the single-particle Hilbert space,
instead of one. So, let us define the dimensionality $d(N)$
of the Hilbert space of states available to an additional particle
given that there are already $N$ particles in the system. (Some
high-energy cutoff is needed, of course, to make this finite.)
The `exclusion statistics parameter' $g$ is, then defined as
\cite{HES}
\be
g = - \frac{\Delta d}{\Delta N}
\label{defg}
\ee
If this parameter is independent of $N$, or becomes a constant for
high enough $N$, then we say that the system obeys exclusion statistics
of order $g$. Clearly $g=0,1$ corresponds to bosons and fermions,
respectively. From the discussion of the CSM model we would conclude
that it obeys exclusion statistics with $g = \ell$.

Note that, in principle, this definition applies to systems of
arbitrary dimensionality. The fact that only one-dimensional exclusion
statistics systems have been as yet identified presumably points to 
something essentially one-dimensional in this definition. Note also
that $g$ can be fractional: $d(N)$ could be either ill-defined (as,
in fact, in the CSM model) so that it can assume ``approximate''
fractional values, or it could depend on $N$ in a way that $g$
becomes constant only for large enough $\Delta N$, like, e.g.,
$d(N) = [N/2]$ with $[ \, ]$ the integer part, giving $g=1/2$.

\subsection{Semiclassics -- Heuristics}

Before examining the consequences of (\ref{defg}) let us make
some heuristic semiclassical arguments about phase space volume
and exclusion to give more substance to it. Please view the
following discussion simply as additional motivation -do not take
it seriously!

Semiclassically, the number of states of a system is given by the
volume of its phase space in units of $h$ for each pair of canonical
variables. Let us consider for simplicity the minimal space $(q,p)$.
It could correspond, for instance, to the relative coordinate and
momentum of two particles on the line. The standard lagrangian would
look like
\be
L = p {\dot q} - H(p,q)
\ee

If we want the presence of the one particle to `knock out' $g$ states
out of the Hilbert space, we should include some term in the
classical action that, effectively, reduces the volume (in this 
case, area) of phase space by $gh$. This area is given as
\be
A = \int_D dp dq = \oint_{\partial D} p dq
\ee
where $V$ is a domain in phase space and $\partial D$ its boundary.
If the domain does not include the point $q=p=0$ the area should be
the standard one, since we are talking about a region of phase space
where the particles are apart. If, on the other hand, $q=p=0$ {\it is}
included, the particles are together and the area should diminish 
by $gh=2\pi g$ (taking $\hbar=1$). 

This reminds us of the Aharonov-Bohm effect. There must be a 
`Dirac string' piercing the point $(0,0)$ in phase space giving this
extra contribution when circled. So we must add to the action a term
\be
\S_g = -\lambda g \int \frac{pdq - qdp}{p^2 + \lambda^2 q^2} 
= -g\int d \left( {\rm atan} \frac{\lambda q}{p} \right)
\ee
where $\lambda$ is any positive constant. This
amounts to adding to the lagrangian the extra term
\be
L_g = -\lambda g\frac{p{\dot q} -q{\dot q}}{p^2 + \lambda^2 q^2}
\ee
This extra contribution is a total time derivative (a topological term)
so it will not change the equations of motion. It is expected, though,
that it will change the quantum mechanical states as described above.

Let us consider the above in the specific example of two particles
in an external harmonic oscillator potential. After separating the
center of mass phase space, the lagrangian for the relative part becomes
\be
L_g = p{\dot q} -\omega g\frac{p {\dot q} -q {\dot q}}{p^2 + \omega^2 q^2}
-\half \omega^2 (p^2 + q^2)
\ee
For later convenience we chose $\lambda = \omega$ in the $g$-term.
The effect of this term is to shift the Poisson brackets between
$p$ and $q$. We will follow the simple approach to canonical quantization
of defining new `polar' variables $\rho, \theta$ as
\be
p + i \omega q = {\sqrt \omega} \rho e^{i\theta}
\ee
In terms of these the lagrangian becomes
\be
L = ( \half \rho^2 -g) {\dot \theta} - \half \omega \rho^2
\ee
so the canonical momentum of $\theta$ is
\be
\pi_\theta = \half \rho^2 -g
\ee
and the hamiltonian becomes
\be
H = \omega ( \pi_\theta + g)
\ee
Quantum mechanically the operator $\theta$ is not quite well defined.
In the absence of $g$, it would correspond to the `phase' of the
annihilation operator which cannot be a well-defined hermitian operator:
$e^{i\theta}$ should be unitary, yet it decreases the eigenvalue
of $\rho^2 = a^\dagger a$ by one unit, which is impossible for the
ground state. Nevertheless, we will proceed qualitatively and see
what we get in this case.

$\theta$ being a phase, its momentum $\pi_\theta$ should be quantized
to integer values. Since the particles are identical, the change of
relative coordinates $p \to -p$, $q \to -q$ should also be a symmetry.
This corresponds to $\theta \to \theta +\pi$. Choosing the wavefunction
to transform as $\pm$ itself under this shift, which amounts to
quantizing the particles as bosons or fermions, further restricts the
values of $\pi_\theta$ to even or odd integers. Finally, since
$\rho^2$ is a positive definite operator, we have
\be
\half \rho^2 = \pi_\theta + g \ge 0
\ee
(clearly the problems with the definition of $\theta$ are hidden
in the above constraint).

Choosing the even eigenvalues for $\pi_\theta$ (bosonic case), and
for $g$ not greater that $2$, the spectrum of the relative hamiltonian
becomes
\be
E = \omega (2n +g)
\ee
for $n$ a nonnegative integer. This is the excitation spectrum of the
Calogero model! To see this, add the center of mass oscillator energy
$E_{cm} = \omega m$ for $m$ a nonnegative integer and define
\be
n_1 = n ~,~~~ n_2 = m+n+g
\ee
Then the full spectrum becomes
\be
E = \omega ( n_1 + n_2 ) ~~{\rm with}~~ n_1 \le n_2 -g
\ee
that is, the excitations of the Calogero model in terms of the
quasiexcitation numbers $n_{1,2}$ satisfying the `least distance
$\ell$' constraint for $\ell = g$.

This will serve as enough motivation that the idea of phase space
exclusion should be related to eigenvalue repulsion {\`a la} CSM
model. The questions about the ground state energy, $g>2$ etc. that
are left hanging are simply set aside -this model is not treated
seriously here. If anyone is interested, of course, they are welcome
to polish it!

\subsection{Exclusion statistical mechanics}

We will implement exclusion statistics in terms of the possible
quantum numbers of $N$ particles placed in $K$ single-particle
states. Arranging the $K$ states in a linear fashion and implementing
the `least distance $g$' constraint for the particles, for integer
$g$ we get \cite{HES,WES}
\be
D(K,N) = \frac{[K-(g-1)(N-1)]!}{N! [K-g(N-1)-1]!}
\ee
possible combinations for allowed values for the particle quantum
numbers (to prove it is a simple combinatorial matter). Extrapolated
to arbitrary (fractional) $g$, this defines the state multiplicity
of $N$ particles placed in $K$ states. Clearly $g=0,1$ reproduces
the Bose and Fermi result.

This can be used to derive the statistical mechanical properties of
a grand ensemble of such particles. Considering all $K$ states to
be at nearby energy $\epsilon$ and maximizing the Gibbs factor
\be
D(K,N) e^{\beta (\mu - \epsilon)N} = maximum
\ee
in terms of $N$ for fixed temperature $T=1/k \beta$ and chemical
potential $\mu$ we obtain for the distribution $n = N/K$ for large
$N,K$ \cite{Isa,WES,Raj}
\be
n = \frac{w}{1+(g-1)w} ~,~~~ \frac{w}{(1-w)^g} = e^{\beta (\mu
- \epsilon)}
\label{excln}
\ee
The expression for $n$ cannot be found analytically except for a
few values of $g$.

It is obvious from the above that fractional statistics particles
are not `independent', in the sense that one cannot derive their
statistical mechanics by considering each single-particle state
as an independent system to be filled by particles. (Operator 
constructions that incorporate some of the features of
exclusion statistics do exist \cite{KN} but they further require
specific interactions, else they produce Gentile statistics.) 
There is no
microscopic formulation: we need to start from $K$ states and take
the limit of large $K$. Further, it is not clear how we could
implement exclusion statistics for an interacting system, where
we cannot use single-particle states as a convenient basis.
In the following we will sketch how the above difficulties
can be overcome \cite{PolES}.

The starting point will be the grand partition function for
particles in $K$ states of like energy
\be
Z (K,z) = \sum_{N=0}^\infty D(K,N) z^N
\ee
where $z=\exp\beta(\mu-\epsilon)$ is the fugacity.
$Z(K,z)$ should be extensive, which means that, for large $K$
it should become the $K$-th power of a function of $z$
\be
\lim_{K \to \infty} \frac{1}{K} \ln Z(K,x) = \lim_{K \to \infty}
\frac{1}{K} \ln\sum_{N=0}^\infty D(K,N) z^N 
= \ln \sum_n P_n z^n
\ee
In the above we introduced the quantities $P_n$ which play a role
analogous to the allowed states of occupation of a single level
in the cases of fermions and bosons: $P_n = 1$ for bosons, while
$P_{0,1} =1$, $P_{n>1} =0$ for fermions. We can call them
`a priori probabilities of occupation' \cite{PolES} or 
`fractional dimensionality of states' \cite{IIG} according to taste.

That $P_n$ exist should be guaranteed from the proper extensive 
behavior of the exclusion statistical mechanics. The problem is to
calculate them. For this, we make a technical trick: to derive the
multiplicity of states $D(K,N)$ above we placed the $K$ states on
a line. Let's place them on a circle instead, and implement the
`minimum distance $g$' rule there. That shouldn't influence the
statistical behavior at large $K$. 
This modifies the combinatorics into
\be
{\tilde D} (K,N) = \frac{K [K-(g-1)N -1]!}{N! (K-gN)!}
\ee
We can check that ${\tilde D} (K,N)$ leads to the same statistical
distribution $n$ (\ref{excln}) as $D(K,N)$.
We now notice that $Z(K,z)$ defined in terms of
${\tilde D} (K,N)$ becomes an exact $K$-th power for all $K$.
We can, then, calculate $P_n$ from ${\tilde D} (1,n)$. The
result is
\be
P_n = \prod_{m=2}^n \Bigl( 1- {gn\over m}\Bigr)
\ee
Unless $g=0,1$, $P_n$ are fractional and always become negative for
some values of $n$ \cite{NW}. So their interpretation as probabilities
or space dimensions must be taken with a grain of salt. They
are, at any rate, useful tools for describing these systems.

The single-level grand partition function $Z(z)$ can be shown 
to satisfy
\be
Z^g - Z^{g-1} = x
\ee
from which the corresponding relation (\ref{excln}) for
$n = z \partial_z \ln Z$ follows.

Remember that the CSM system enjoyed a particle-hole 
coupling-inverting duality. Since exclusion statistics are directly
extracted from the properties of these systems, we anticipate
a similar duality here \cite{NW,PolEX}. Indeed, we can show that 
the grand partition function satisfies
\be
\frac{1}{Z (g,x^{-g})} + \frac{1}{Z (g^{-1} ,x)} = 1
\label{dualZ}
\ee
Note that, interestingly, a similar relation holds for Gentile
statistics of maximum occupancy $p$. The grand partition 
function obviously is
\be
Z(p,z) = 1+z+z^2 \dots +z^p = \frac{z^{p+1} -1}{z-1}
\ee
Identifying the maximum occupancy $p$ of Gentile 
statistics states with the parameter $1/g$, the above expression
satisfies (\ref{dualZ}) although $Z(p,z)$ is not a priori defined
for fractional $p$.

{}From the distribution function $n(g,z)$ (\ref{excln}) we can
derive in the standard way the low-temperature Sommerfeld expansion
for the energy
\be
E(T) = \int d\epsilon \epsilon \rho(\epsilon) n(g,\epsilon) =
E(0) + \sum_{n=0}^\infty C_n (g) T^{n+1} E_n
\ee
where $\rho(\epsilon)$ is the density of states and $\mu$ 
is fixed through
\be
N = \int d\epsilon \rho(\epsilon) n(g,\epsilon) 
\ee
$E_n$ are $T$- and $g$-independent energy integrals.
The coefficients $C_n$ essentially determine the low-temperature heat
capacity of the system. The are calculated as \cite{AIMP}
\begin{eqnarray}
C_0 &=& 0  ~~~~~~ {\rm (third~law~of~thermodynamics)}\\
C_1 &=& \frac{\pi^2}{6} ~~~~{\rm (same~as~in~conformal~field~theory)}\\
C_2 &=& 2 \zeta(3) (1-g) ~~etc.
\end{eqnarray}
where $\zeta(x)$ is the Riemann zeta function.
In general, $C_n$ is a polynomial of degree $n-1$ in $g$.

The grand potential $G=-kT \ln Z$ can be expressed in terms of
cluster coefficients $w_n$:
\be
\ln \sum_{n=0}^\infty P_n z^n = \sum_{n=1}^\infty {w_n \over n} z^n
\label{clstw}
\ee
We find for $w_n$:
\be
w_n = \prod_{m=1}^{n-1} \Bigl( 1 - {gn \over m} \Bigr)
\ee
which is remarkably similar to $P_n$ (except for the range of $m$).
Similar results were obtained in \cite{dVO} for the case of anyons
in a strong magnetic field. This is not surprising: the lowest
Landau level becomes, essentially, a two-dimensional phase space,
so this is yet another realization of fractional statistics in an
effectively one-dimensional space.

Note that the connection between $P_n$ and $w_n$ is the same as
the one between `disconnected' and `connected' diagrams in field
theory. This, in view also of the discussion of the first section,
will lead us to a path-integral representation for the partition 
function of exclusion statistics particles in an arbitrary 
external potential. 

\subsection{Exclusion statistics path integral}

We start from the usual euclidean path integral
with periodic time $\beta$ for $N$ particles with action the sum of
$N$ one-particle actions; we further sum over all particle numbers 
$N$ with chemical potential weights $e^{\beta \mu N} /N!$ to obtain 
the grand partition function. Since the particles are identical we
include a symmetry factor of $1/N!$, but we must
also sum over paths where particles have exchanged final positions.
Thus the path integral for each $N$ decomposes into sectors 
labeled by the elements of the permutation group $Perm(N)$.
(See the discussion in section 1.) Such permuted sectors will be 
summed with appropriate extra weighting factors, to be determined
in the sequel.

By the usual argument, the Gibbs grand potential (the logarithm of
the grand partition function) will be given by the sum
of all connected path integrals. It is obvious that these are
the ones where the final positions of the particles are a
cyclic permutation of the original ones, since these are the
only elements of $S_N$ that cannot be written as a product
of commuting elements. Clearly all diagrams corresponding to 
different cyclic permutations are equal. There are $(N-1)!$ such
permutations. So, overall, this diagram will carry a factor of 
$((N-1)!/N! = 1/N$. This is nothing but the `symmetry factor,'
familiar from Feynman diagrams, corresponding to cyclic relabelings
of particle coordinates (compare, already, with the factors $1/n$
included in (\ref{clstw})).

Cyclic permutation diagrams being connected, they only contain
one `thread' of particle worldline. They really correspond to one 
particle wrapping $N$ times
around euclidean time and thus will reproduce the single-particle
partition function with temperature parameter $N\beta$. So, in the
grand partition function they will contribute the terms proportional
to $e^{-N\beta \epsilon} e^{\beta \mu N} = z^N$ (the first factor
is the single-particle Boltzmann factor for temperature parameter
$N \beta$ and the second is the chemical potential factor).

But these are the terms appearing in (\ref{clstw}) for each energy 
level! The extra weighting factors that must be included for
these diagrams in order to reproduce exclusion statistics are,
then, identified to the cluster coefficients $w_N$. For $g=0,1$
we have $w_N =1$ and $w_N = (-)^{N-1}$, respectively, which is,
indeed, the correct factor implied by $\chi_R (P)$ as discussed
in section 1, where $R$ is the trivial (bosons) or the antisymmetric
(fermions) irrep of $S_N$ and $P$ a cyclic permutation.

In conclusion, if we weight these configurations
with the extra factors $w_N$ we will
reproduce the grand potential of a distribution of $g$-ons on the
energy levels of the one-body problem, that is \cite{PolES}
\be
\Omega (\beta,\mu) = \sum_{N=1}^\infty e^{\mu N} \frac{w_{_N}}{N}
\int e^{-\sum_{n=1}^N S_E [ x_n (t_n ) ]}
\prod_{n=1}^N Dx_n (t_n ) 
\label{PIfree}
\ee
where $S_E$ is the one-particle euclidean action and the paths
obey the boundary conditions 
\be
x_n (\beta) = x_{n+1} (0) ~,~~~ x_N (\beta) = x_0 (0)
\ee
($x$ can be in arbitrary dimensions.)

By exponentiating, the full $N$-body partition function will be the
path integral over all disconnected components, with appropriate
symmetry factors and a factor of $w_n$ for each connected
$n$-particle component. More directly, we can omit the symmetry
factors, sum explicitly over all equivalent permutations in each
sector and divide by $N!$:
\be
{\cal Z}(\beta,\mu) = \sum_{N=1}^\infty \frac{e^{\mu N}}{N!}
\sum_P W_P \int e^{-\sum_{n=1}^N S_E [ x_n (t_n ) ]}
\prod_{n=1}^N Dx_n (t_n ) 
\label{PIexp}
\ee
where the paths obey the boundary condition
\be
\{x_1 (\beta) , \dots x_N (\beta) \} = P \{ x_1 (0) , \
\dots x_N (0) \}
\ee
The weighting factor $W_P$ for each permutation depends only on the
conjugacy class of the permutation and is calculated through $w_n$ as
\be
W_P = w_{n_1} \cdots w_{n_k}
\ee
where $n_1 ,\dots n_k$ are the cycles of $P$. 

It is clear that the above path integral is not unitary, since
the weights $w_n$ are not phases and they do not provide true 
representations of the permutation group (unlike the $g=0,1$ cases).
This does not matter: it is simply used as a tool to derive
the statistical mechanics of exclusion statistics. We are not
going to calculate and propagators or other processes with it.

Once we have the path integral (\ref{PIexp}) we can easily extend 
the notion of exclusion statistics
to interacting particles: we simply replace the action 
$\sum_n S_E [x_n ]$ by the full interacting $N$-particle action,
thus circumventing all difficulties with combinatorial formulae.
In the interacting case one has to work with the
full grand partition function, rather than the grand potential,
since topologically disconnected diagrams are still dynamically
connected through the interactions and do not factorize.

In conclusion, we have a path-integral way of defining exclusion
statistics for interacting systems through appropriate symmetry
factors included in each permutation sector (analogous
to $S(P)$ of section 1), calculated via (\ref{clstw}). There is,
yet, no application of this procedure to an interesting interacting
situation.

\subsection{Is this the only `exclusion' statistics?}

After giving so much motivation for it, the question seems
almost offensive! But let us examine it for the sake of ensuring
that we have the full picture.

Exclusion statistics in terms of `repulsion rules' for the quantum
numbers is, certainly, as we defined it. But we started from 
(\ref{defg}) which defines statistics in terms of the Hilbert space
dimension
of an additional particle in the system. Consider, then, a chunk of
$K$ levels in the CSM model ($K$ quasimomentum values in the Sutherland
model or $K$ quasiexcitation levels for the harmonic Calogero model).
Place $N$ particles there, and see what space is left for an 
additional particle.

If the $N$ particles are packed `densely' (like, e.g., in the Fermi
sea-like ground state) they certainly take up a piece of approximately
$\ell N$ spaces. So the levels left for an extra particle are 
$d = K-\ell N$ and we get $g = -\Delta d/\Delta N = \ell$ as
expected.

What if, however, the $N$ particles are `sparse' in $K$; that is,
if the distances between them are all bigger than, say, $2\ell$?
Then each particle makes unavailable $\ell$ states {\it either way}
around it, so, overall, the available space for the extra particle has
diminished by $(2\ell -1)N$. In this situation we would get 
$g = 2\ell -1$. Clearly this definition implies different statistics
for different situations (dense or dilute).

Can we define another statistics that matches closer the Hilbert
space definition? Clearly any choice of combinatorial formula
for $D(K,N)$ that has the right extensive properties is an
alternative definition of {\it some} statistics. Alternatively,
each choice of $P_n$ or $w_n$ amounts to some statistics.
Let's make the simplest choice:
\be
w_n = (-\alpha)^{n-1}
\ee
which, in the path integral, corresponds to one factor of $-\alpha$ 
for each unavoidable particle crossing. This leads to the statistical
distribution ${\bar n}$
\be
{\bar n} = {1 \over e^{(\epsilon - \mu)\beta} + \alpha}
\ee
which was analyzed in \cite{ANS} as the simplest imaginable 
generalization of the Fermi and Bose distribution. 
The combinatorial formula for $D(K,N)$
for the above $\alpha$-statistics is
\be
D = \alpha^N {({K\over \alpha })! \over N! ({K\over \alpha} -N)!}
= {K (K-\alpha ) (K-2\alpha ) \cdots (K-(N-1)\alpha )\over N!}
\ee
This can be thought as a realization of the exclusion statistics 
Hilbert space idea: the first particle put in the system has $K$
states to choose, the next has $K-\alpha$ due to the presence of
the previous one an so on, and dividing by $N!$ avoids overcounting.
Fermions and bosons correspond to $\alpha=1$ and $\alpha=-1$
respectively. $\alpha=0$ corresponds to Boltzmann `classical'
statistics. This should also be clear from the path integral: no 
configurations where particles have exchanged positions are
allowed, since their weighting factor is zero, but factors of 
$1/N!$ are still included. The corresponding single-level 
$P_n$ are
\be
P_n = \prod_{m=1}^{n-1} {1-m\alpha \over 1+m}
\ee
For $\alpha=1/p$ with $p$ integer (a fraction of a fermion),
the above `probabilities' are all positive for $n$ up to $p$
and vanish beyond that. For $\alpha<0$ all probabilities are
positive and nonzero. Thus, the above system has a bosonic
($\alpha <0$) and a fermionic ($\alpha>0$) sector, with
Boltzmann statistics as the separator. It is a plausible
alternative definition of exclusion statistics,
and has many appealing features, not shared by the standard
exclusion statistics, such as positive probabilities, 
a maximum single-level occupancy in accordance with the fraction 
of a fermion that $\alpha$ represents, and analytic expressions 
for all thermodynamic quantities. 

This should demonstrate that the route to alternative definitions,
with nice features too, is open. The real issue is whether a
dynamical system realizes these other statistics, and, at this 
point, this is not clear.

\section{Epilogue}

The topic of statistics will not cease fascinating at least
a few physicists working in various fields. It challenges our
understanding of the fundamentals of quantum mechanics and matter,
expands the mathematical tools of the trade, promises new
results in specific systems and is fun to think about.

It is anyone's guess whether considerations coming purely
out of statistics will produce a (major or minor) breakthrough in
any physics problem. A safer prediction, however, is that the
idea of generalized statistics will re-emerge in different contexts
as we strive to understand new physical systems and develop useful
frames of mind. After all, statistics has barely been touched for
objects such as strings and membranes, which are increasingly the
entities of theoretical choice for the fundamental constituents
of the universe. The future promises to be fun!

\end{document}